\AddToHookNext{shipout/foreground}{%
  \put(0pt,-\paperheight){%
    \raisebox{0.5in}[0pt][0pt]{%
      \makebox[\paperwidth][c]{%
        \parbox[b]{\textwidth}{%
          \normalfont\footnotesize\centering
          This work has been submitted to the IEEE for possible publication.\\
          Copyright may be transferred without notice, after which this version
          may no longer be accessible.%
        }%
      }%
    }%
  }%
}

\documentclass[10pt,conference,letterpaper]{IEEEtran}
\IEEEoverridecommandlockouts

\usepackage[T1]{fontenc}
\usepackage{amsmath,amssymb,amsfonts}
\usepackage{amsthm}
\usepackage{mathtools}
\usepackage{dsfont}
\usepackage{array}
\usepackage{multirow}
\usepackage{booktabs}
\usepackage{cite}
\usepackage{etoolbox}
\usepackage{graphicx}
\usepackage{algorithm}
\usepackage{algpseudocode}
\usepackage{enumitem}
\usepackage{xspace}
\usepackage{acro}
\usepackage{xurl}
\usepackage{subfiles}
\usepackage[hidelinks]{hyperref}

\graphicspath{{figures/}{../figures/}}

\newif\ifcitationseen
\pretocmd{\cite}{\global\citationseentrue}{}{}
\pretocmd{\nocite}{\global\citationseentrue}{}{}

\newcommand{\standalonereferences}{%
  \ifSubfilesClassLoaded{%
    \ifcitationseen
      \bibliographystyle{IEEEtran}%
      \bibliography{references}%
    \fi
  }{}%
}

\newcommand{\reqsla}{\textit{(R1)}\xspace}
\newcommand{\reqcost}{\textit{(R2)}\xspace}
\newcolumntype{P}[1]{>{\raggedright\arraybackslash}p{#1}}

\theoremstyle{definition}
\newtheorem{definition}{Definition}
\newtheorem{assumption}{Assumption}
\newtheorem{remark}{Remark}
\newtheorem*{remark*}{Remark}

\theoremstyle{plain}
\newtheorem{lemma}{Lemma}
\newtheorem{theorem}{Theorem}
\newtheorem{proposition}{Proposition}

\acsetup{first-style = long-short}

\DeclareAcronym{embb}{short = eMBB, long = enhanced mobile broadband}
\DeclareAcronym{urllc}{short = URLLC, long = ultra-reliable and low-latency communications}
\DeclareAcronym{mmtc}{short = mMTC, long = massive machine-type communications}
\DeclareAcronym{ran}{short = RAN, long = radio access network}
\DeclareAcronym{gnb}{short = gNB, long = next-generation NodeB}
\DeclareAcronym{tn}{short = TN, long = transport network}
\DeclareAcronym{cn}{short = CN, long = core network}
\DeclareAcronym{ns}{short = NS, long = network slicing}
\DeclareAcronym{sla}{short = SLA, long = service-level agreement}
\DeclareAcronym{nst}{short = NST, long = network slice template}
\DeclareAcronym{gst}{short = GST, long = Generic Network Slice Template}
\DeclareAcronym{nest}{short = NEST, long = Network Slice Type}
\DeclareAcronym{gsma}{short = GSMA, long = GSM Association}
\DeclareAcronym{ietf}{short = IETF, long = Internet Engineering Task Force}
\DeclareAcronym{yang}{short = YANG, long = Yet Another Next Generation}
\DeclareAcronym{mec}{short = MEC, long = multi-access edge computing}
\DeclareAcronym{iot}{short = IoT, long = Internet of Things}
\DeclareAcronym{rb}{short = RB, long = resource block}
\DeclareAcronym{upf}{short = UPF, long = user plane function}
\DeclareAcronym{ue}{short = UE, long = user equipment}
\DeclareAcronym{knscp}{short = $k$-NSCP, long = $k$-network slice catalog problem}
\DeclareAcronym{milp}{short = MILP, long = mixed-integer linear program}
\DeclareAcronym{minlp}{short = MINLP, long = mixed-integer nonlinear program}
\DeclareAcronym{saa}{short = SAA, long = sample average approximation}
\DeclareAcronym{lbbd}{short = LBBD, long = logic-based Benders decomposition}
\DeclareAcronym{iid}{short = i.i.d., long = independent and identically distributed}

\title{Demand-Aware Network Slice Catalog Design}

\author{%
  \IEEEauthorblockN{%
    Masaki Kobayashi\IEEEauthorrefmark{1},
    Akito Suzuki\IEEEauthorrefmark{1},
    Ryoichi Kawahara\IEEEauthorrefmark{2}, and
    Masahiro Kobayashi\IEEEauthorrefmark{1}}
  \IEEEauthorblockA{\IEEEauthorrefmark{1}%
    Network Service Systems Laboratories, NTT, Inc.\\
    3--9--11 Midori-cho, Musashino-shi, Tokyo 180--8585, Japan\\
    Email: \{kobayashi.m, akito.suzuki, masahiropk.kobayashi\}@ntt.com
  }
  \IEEEauthorblockA{\IEEEauthorrefmark{2}%
    Faculty of Information Networking for Innovation and Design, Toyo University\\
    Kita-ku, Tokyo 115--8650, Japan\\
    Email: ryoichi.kawahara@iniad.org
  }
}

\providecommand{\CurrentMSTenThree}{0.29}
\providecommand{\CurrentMSTenFive}{14.39}
\providecommand{\CurrentSmallInstances}{20}
\providecommand{\CurrentSmallLtwoHits}{4}
\providecommand{\CurrentSmallLtwoError}{5.92}
\providecommand{\CurrentSmallHellingerHits}{0}
\providecommand{\CurrentSmallHellingerError}{31.47}
\providecommand{\CurrentRingMSThree}{0.28}
\providecommand{\CurrentRingMSFive}{14.54}
\providecommand{\RefinedAnyTrainStart}{2.164}
\providecommand{\RefinedAnyTrainEnd}{1.810}
\providecommand{\RefinedAnyTestStart}{2.204}
\providecommand{\RefinedAnyTestEnd}{1.862}
\providecommand{\RefinedAnyEndCalls}{185}
\providecommand{\RefinedAnyTimeLimited}{3}
\providecommand{\RefinedAnyReachedCap}{7}
\providecommand{\RefinedSmallMinCalls}{13}
\providecommand{\RefinedSmallMaxCalls}{35}
\providecommand{\RefinedMainMinGap}{26.94}
\providecommand{\RefinedMainMaxGap}{32.06}

\begin{document}

\maketitle

\begin{abstract}
In network slicing, a user requests a slice by specifying \ac{sla}
requirements, such as latency and bandwidth, in a network slice template, and
is charged a provisioning cost. Because preparing
a slice for new requirements involves multi-domain design and verification,
which delays provisioning, the slice operator can predesign a finite set of
validated templates, called a network slice catalog, and serve each request
with a catalog template satisfying all its requirements.
A catalog designed without regard to user demand may serve a request only with
a template far stricter than requested, which allocates excess resources and
increases the provisioning cost, or not at all, requiring bespoke design and
validation at an additional cost to the user.
We therefore aim to design a catalog of $k$ templates that minimizes the
expected provisioning cost under user demand.
To this end, we propose (i) the \ac{knscp}, which formalizes this design task,
and (ii) a method that solves it exactly.
The formulation expresses \ac{sla} satisfaction through a dominance relation:
one template dominates another if every slice satisfying the former
also satisfies the latter.
Under the standardized hierarchical slice management architecture, the slice
operator can obtain provisioning costs, but the details of the underlying
resource allocation algorithm are not disclosed.
Without requiring access to these algorithmic details, our method provably
finds an optimal catalog using only finitely many cost evaluations,
even when the template space is infinite.
In experiments on tree and ring networks, our method reduced the provisioning
cost by up to \CurrentMSTenFive\% and
\CurrentRingMSFive\%, respectively, relative to the strongest clustering-based
baseline.
\end{abstract}

\begin{IEEEkeywords}
5G, network slicing, mathematical optimization
\end{IEEEkeywords}

\acresetall

\section{Introduction}
\label{sec:introduction}

\Ac{ns} has attracted attention as a means of supporting the
diverse requirements of services such as \ac{urllc} and \ac{embb}.
The interface through which a user requests a slice is standardized in terms of
a tuple of \ac{sla} requirements that the slice is expected to satisfy~\cite{generic_slice_template};
we refer to such a tuple as a \textit{network slice template}.
To request a slice, the user specifies the desired requirements in a
user-specified template. For example, a user requesting a slice with
a latency of at most $10\,\mathrm{ms}$ and a bandwidth of at least
$100\,\mathrm{Mbps}$ specifies the template
$(10\,\mathrm{ms}, 100\,\mathrm{Mbps})$.
Resources in the 5G network are then allocated so that the requirements in the
specified template are satisfied.
The user is charged for provisioning the slice; we call
this charge the \emph{provisioning cost}.

\begin{figure}[t]
    \centering
    \includegraphics[width=0.77\linewidth]{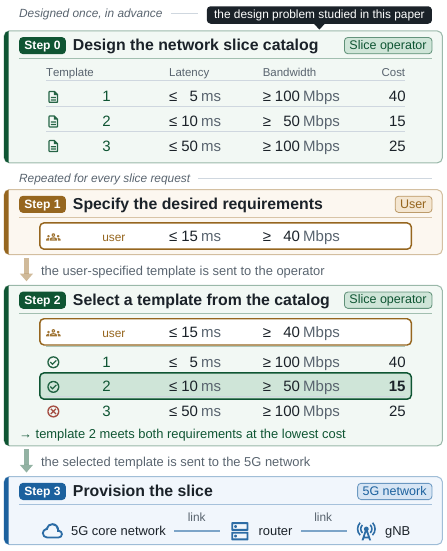}
    \caption{Four steps of catalog-based slice provisioning: Step 0) the slice operator designs a catalog; Step 1) a user specifies the desired requirements in a user-specified template; Step 2) the operator selects a catalog template; and Step 3) the 5G network allocates resources according to the selected template.}
    \label{fig:catalog_design_policy}
\end{figure}

Preparing a new slice involves tasks spanning
the radio access, transport, and core network domains, such as designing the
slice, on-boarding and verifying network functions, and checking
feasibility~\cite{3gpp28530}. These tasks may require operator intervention,
hindering fast provisioning~\cite{samsung2020network_slicing}.

To avoid repeating this time-consuming preparation for every request and
thereby enable fast, automated provisioning, the slice operator can prepare
and validate a finite set of slice templates in advance. When a user
submits a template specifying the desired requirements, the operator selects
from this set a template that satisfies the requirements in the user's
template and provisions the slice accordingly.
This operating model is illustrated in the 3GPP provisioning
specification~\cite{3gpp28531}.
It has also been identified as a promising way to automate and accelerate
slice provisioning~\cite{samsung2020network_slicing,5gamericas2022commercializing} and
is already supported by commercial software~\cite{cisco_cwnc7swg,vmware_tca_slice_catalog_32}.
We call the set of predesigned templates a \emph{network slice catalog} (Fig.~\ref{fig:catalog_design_policy}).

A poorly designed catalog can increase the provisioning cost in two ways.
First, consider the catalog in Fig.~\ref{fig:catalog_design_policy} and
the user-specified template $(40\,\mathrm{ms}, 50\,\mathrm{Mbps})$.
Among the catalog templates satisfying the requested requirements, $(10\, \mathrm{ms}, 50\,\mathrm{Mbps})$ has the lowest provisioning cost.
The operator must therefore select this template and allocate additional
resources to satisfy the stricter latency requirement, increasing the
provisioning cost.
Second, consider the user-specified template
$(5\,\mathrm{ms}, 120\,\mathrm{Mbps})$, for which no catalog template
satisfies the requested requirements. The operator must then prepare and
validate a bespoke template, incurring additional operational cost.
We assume that this additional cost is passed on to the user, thereby
increasing the provisioning cost.
Catalog design must therefore account for the user demand.

\textbf{Research Objective:} We aim to reduce provisioning cost through
catalog design that accounts for user demand. Specifically, we design a
catalog of $k$ templates that minimizes the expected provisioning cost under
the user demand distribution, using the user-specified templates submitted
in past requests as samples.
The catalog is designed offline from these samples and then deployed
to provision future requests.

\textbf{Related Work:}
Designing a catalog from such samples can be viewed as a clustering
problem: the samples are partitioned into at most
$k$ groups, and a catalog template is designed for each group.
Such template clustering methods have been explored~\cite{otieno2024review}, including methods based on
$k$-means~\cite{min2024} (Sec.~\ref{sec:related_work}).
However, standard clustering formulations such as $k$-means do not incorporate
the following two requirements of catalog design:
\reqsla\ \textit{\ac{sla} satisfaction:} each request served from the catalog
must be assigned to a template satisfying all requested \ac{sla} requirements; and
\reqcost\ \textit{Provisioning-cost objective:} catalog design must minimize
provisioning cost.
For example, if $k$-means groups two bandwidth requests of
$100\,\mathrm{Mbps}$ and $200\,\mathrm{Mbps}$ into the same cluster,
the centroid specifies $150\,\mathrm{Mbps}$ and fails to satisfy
the latter request \reqsla.
Moreover, $k$-means uses an objective based on Euclidean distance
and does not aim to minimize provisioning cost \reqcost.

\textbf{Our Method:}
We propose (i) the \acf{knscp}, a stochastic program that formulates catalog
design with \reqsla\ \textit{\ac{sla} satisfaction} and
\reqcost\ \textit{Provisioning-cost objective}, and
(ii) an exact algorithm for its \ac{saa} problem.
\ac{saa}, a standard approach to stochastic
programming~\cite{kleywegt2002saa}, replaces the expected provisioning cost
with its average over the samples; under standard conditions, the optimal
value of the \ac{saa} problem converges to that of the \ac{knscp} as the
number of samples grows.
Together, the formulation and algorithm enable optimal catalog design
from the sample templates.

\textit{(i) Formulation:}
We express \reqsla\ \textit{\ac{sla} satisfaction} through a dominance
relation between templates: template $s_2$ dominates template $s_1$ if every slice
satisfying all \ac{sla} requirements specified by $s_2$ also satisfies
all those specified by $s_1$.
Using this relation, we incorporate
\reqsla\ \textit{\ac{sla} satisfaction} into the \ac{knscp} formulation
(Sec.~\ref{sec:problem_definition}).

\textit{(ii) Algorithm:}
Solving the \ac{saa} problem for the \ac{knscp} is challenging for two
reasons.
(a) \textit{Black-box costs:}
{The slice operator can query the provisioning cost of a given
template, but the details of the underlying resource allocation algorithm
are not disclosed.}
This black-box access reflects the hierarchical architecture standardized
for slice management~\cite{etsi_zsm_003}.
Without an explicit functional form for these costs, the \ac{saa} problem
cannot be passed directly to a standard optimization solver.
(b) \textit{Infinite template space:} The template space may be
infinite, so the costs of all templates cannot be evaluated; an optimal
catalog must be found from finitely many cost evaluations.

To address these challenges, we exploit the monotonicity of provisioning cost
induced by template dominance: \textit{if $s_2$ dominates $s_1$, the
provisioning cost of $s_2$ is never lower than that of $s_1$, because $s_2$
imposes stricter requirements than $s_1$.}
Based on this monotonicity, we develop an iterative algorithm that alternates
between two problems.
(1) The master problem optimizes the catalog under an \textit{optimistic cost
model}, which uses monotonicity to bound unevaluated costs from below by the
evaluated ones.
(2) The subproblem evaluates, through black-box queries, the true costs of the
templates selected by the master problem, and the new evaluations tighten
the optimistic cost model.
Thus, the algorithm requires no explicit functional form of the cost,
addressing (a) \textit{black-box costs}.
Moreover, the algorithm is guaranteed to terminate after finitely many cost
evaluations with a catalog within a prescribed optimality tolerance, even
when the template space is infinite, addressing (b) \textit{infinite
template space} (Theorem~\ref{thm:termination}).

The contributions of this paper are summarized as follows.
\begin{itemize}
    \item We formulate catalog design as the \ac{knscp}, which
      minimizes expected provisioning cost over $k$ predesigned templates
      and incorporates \reqsla\ \textit{\ac{sla} satisfaction} and
      \reqcost\ \textit{Provisioning-cost objective}.

    \item We develop an exact method based on
      the dominance structure of templates and the monotonicity of provisioning cost.
      It combines a finite master \ac{milp} with a closed-form subproblem.

    \item Our method reduced held-out provisioning
      cost by up to \CurrentMSTenFive\% and \CurrentRingMSFive\% on tree and
      ring networks, respectively, relative to the strongest clustering-based
      baseline.
\end{itemize}

\standalonereferences

\section{Related Work}
\label{sec:related_work}

\subsection{Network Slice Template and Catalog Design}
\label{subsec:template_catalog_design}

\subsubsection{Standardization and Industrial Needs}
Reusable templates are part of the standardized operating model for network
slicing. The \ac{gsma} defines a \ac{gst} and its instantiated form, a \ac{nest},
to represent a network slice through generic attributes and their concrete values~\cite{generic_slice_template}.
3GPP illustrates how a provider builds slice products on \acp{nest} and
offers them through a product catalog~\cite{3gpp28531}.
The Network Slice Service \ac{yang} model being standardized within the
\ac{ietf} similarly allows a provider to maintain reusable templates and apply
them to multiple services~\cite{wu2026yang}.
Moreover, catalog-based slice provisioning has been implemented in
practical systems~\cite{cisco_cwnc7swg,vmware_tca_slice_catalog_32},
thereby making catalog optimization valuable for reducing provisioning cost in
practice.

These specifications and commercial systems define \textit{how templates are represented and reused},
but neither the specifications nor the commercial systems determine \textit{which combinations of
\ac{sla} values should compose a finite catalog}.

\subsubsection{Catalog Design Problem}
A substantial body of work studies algorithms for template-based slice
provisioning~\cite{su2019resource_allocation}. By contrast, deciding which templates
should be offered as a finite catalog has received comparatively less
attention.
Within this comparatively limited body of work, researchers have explored
a machine-learning strategy for deriving template attributes~\cite{otieno2023},
discussed the applicability of unsupervised approaches~\cite{otieno2024review}, and proposed a $k$-means-based
method for constructing slice configurations~\cite{min2024}.
These studies suggest clustering as a natural starting point for
designing a catalog.

\subsection{Related Optimization Problems and Methods}
\label{subsec:related_blackbox}

\subsubsection{Clustering and \texorpdfstring{$k$}{k}-Median Problems}
\label{subsubsec:related_clustering}
The \ac{saa} problem for the \ac{knscp} resembles both
$k$-means clustering~\cite{Lloyd1982} and the metric $k$-median
problem~\cite{hakimi1965median}: each selects $k$ representatives
and assigns request samples to them. However, as detailed in
Sec.~\ref{subsubsec:saa}, their distance-based
formulations satisfy neither \reqsla\ \textit{\ac{sla} satisfaction}
nor \reqcost\ \textit{Provisioning-cost objective}.

A closely related problem is continuous container
selection~\cite{nagarajan2015container}, which designs $k$ container
sizes for cloud tasks. It incorporates dominance constraints \reqsla\
and an objective based on resource cost \reqcost, but differs from
our work in the following respects.
\emph{Problem setting:} Container selection uses an explicit $\ell_1$
resource cost and requires every task to be served by a selected
container. In our setting, monotone provisioning costs are available
only through black-box evaluations, and requests may be handled outside
the catalog at a penalty.
\emph{Solution method:} The methods for container selection exploit the
explicit $\ell_1$ cost structure, whereas our method uses cost evaluations
and monotonicity to construct lower bounds.
Because the container selection methods assume $\ell_1$ costs, they
cannot be directly applied to solve the \ac{saa} problem for the
\ac{knscp}.

\subsubsection{Expensive Black-Box Optimization}
\Ac{lbbd} decomposes a problem into a master problem and subproblems
and strengthens the master with valid cuts derived through
problem-specific inference~\cite{hooker2003lbbd}.
Forbes et al.~\cite{forbes2024simulation} derive monotonicity-based cuts
for simulation-based optimization over finite integer decision sets.
Our method can be interpreted as an instance of \ac{lbbd},
as in Forbes et al.
We specialize this approach to catalog design in two respects.
First, we derive an exact \ac{milp} formulation of the master problem
(Sec.~\ref{subsubsec:milp}, Proposition~\ref{prop:milp})
and closed-form optimal solutions to the subproblems
(Sec.~\ref{subsubsec:join}, Lemma~\ref{lem:join}).
Second, we prove that the master attains the strongest lower bound on
the \ac{saa} optimum that can be certified using only the evaluated
costs and monotonicity
(Sec.~\ref{subsubsec:master}, Proposition~\ref{prop:best_bound}).

Bayesian optimization~\cite{baptista2018bocs} also addresses expensive
black-box costs but relies on a probabilistic surrogate model and provides
no deterministic optimality certificate; our method instead exploits
cost monotonicity to certify the gap.

\standalonereferences

\section{Problem Setting: The \texorpdfstring{\acs{knscp}}{k-NSCP}}
\label{sec:problem_definition}

\textbf{Notation.}
Throughout, $\mathbb{R}_{\ge 0}$ denotes the set of non-negative real numbers,
$\mathbb{N}$ the set of integers greater than or equal to one, and
$[a] := \{1,\ldots,a\}$ for $a \in \mathbb{N}$.
For a set $X$ we write $X^{a}$ for the $a$-fold Cartesian product of $X$.
The main symbols used throughout this paper are summarized in
Table~\ref{tab:variables}.

\begin{table}[t]
\caption{Main symbols throughout the paper.}
\label{tab:variables}
\centering
\scriptsize
\begin{tabular}{@{}l@{\hspace{5pt}}P{0.72\linewidth}@{}}
    \toprule
    Symbol & Description\\
    \midrule
    \multicolumn{2}{@{}l}{\textbf{Problem setting}}\\
    \addlinespace[2pt]
    $r_1, \ldots, r_d$ & Requirements on $d \in \mathbb{N}$ attributes\\
    $R_1, \ldots, R_d$ & Sets of values that each attribute may take\\
    $s := (r_1, \ldots, r_d)$ & Network slice template\\
    $S := R_1 \!\times\! \cdots \!\times\! R_d$ & Set of network slice templates\\
    $\operatorname{pr}_j : S \rightarrow R_j$ & Projection extracting the requirement on
      attribute $j$\\
    $\pi \in \mathbb{R}^{p}_{\ge 0}$ & Unit prices of the
      $p \in \mathbb{N}$ resource types\\
    $\Omega(s) \subseteq \mathbb{R}^{p}_{\ge 0}$ & Closed set of resource
      allocations $\xi \in \mathbb{R}^{p}_{\ge 0}$\\
    $\gamma : S \rightarrow \mathbb{R}^{p}_{\ge 0}$ & Black-box function returning a
      resource allocation\\
    $g : S \rightarrow \mathbb{R}_{\ge 0}$ & Provisioning cost attributable to
      resource allocation for a template\\
    $\preceq_j,\ \preceq$ & Total order on the requirement values of
      attribute $j$ and the induced dominance relation on templates\\
    $U \subseteq S$ & Network slice catalog\\
    $k \in \mathbb{N}$ & Size of the catalog $U$\\
    $U_s$ & Catalog templates that satisfy the requirements in $s$\\
    $M$ & Fixed penalty approximating the total provisioning cost
      of a request handled outside the catalog\\
    $f : 2^{S} \times S \rightarrow \mathbb{R}_{\ge 0}$ & Provisioning
      cost under catalog $U$ for a user-specified template $s\in S$\\
    $\mathcal{D}$ & Distribution of user-specified templates\\
    $S_n$ & Set of $n$ sample templates drawn from $\mathcal{D}$\\
    $f_n(U),\ f_n^\star$ & Sample-average cost of catalog $U$ and its
      optimal value\\
    \midrule
    \multicolumn{2}{@{}l}{\textbf{Solution method}}\\
    \addlinespace[2pt]
    $m_\ell$ & $\ell$-th template in the catalog\\
    $x_{i\ell},\ z_i$ & Indicators for assigning sample $s_i$ to template
      $m_\ell$ and handling it outside the catalog\\
    $y=(x,z),\ \mathcal Y$ & Assignment tuple and set of feasible
      assignments\\
    $\Phi_h(m,y),\phi_h(y),\phi_h^\star$ & Catalog cost of $(m,y)$, fixed-assignment
      cost of $y$, and optimal catalog cost under a cost model $h$\\
    $E$ & Set of templates at which $g$ has been evaluated\\
    $\mathrm{LB},\ \mathrm{UB}$ & Current lower and upper bounds on
      $f_n^\star$\\
    $\underline g_E,\ \mathcal H_E$ & Optimistic cost and the set of monotone
      cost models agreeing with $g$ on $E$\\
    $\mathcal R_E$ & Representative templates of the master\\
    $C_\ell(y),\ n_\ell(y)$ & Samples assigned to template $m_\ell$ and their
      number\\
    $\bigvee T$ & Upper template of a non-empty finite template set $T$\\
    $J(y)$ & Upper templates of the nonempty clusters of $y$\\
    \bottomrule
\end{tabular}
\end{table}

\subsection{Network Slice Templates and Provisioning Cost}
\label{subsec:templates_and_provisioning}

\subsubsection{Network Slice Template}
\label{subsubsec:network_slice_template}

We define a template $s$ as a tuple $s := (r_1,\ldots,r_d)$ of requirements
$r_1,\ldots,r_d$ on $d \in \mathbb{N}$ attributes, such as latency and
bandwidth.
For each attribute $j \in [d]$, let $R_j$ denote the \textit{requirement set},
that is, the set of possible requirement values for attribute $j$.
For example, if attribute $j \in [d]$ is latency, then $R_j$ is a subset of
$\mathbb{R}_{\ge 0}$, namely the set of latency values that may be requested.
We define the set $S$ of templates as the Cartesian product of
$R_1,\ldots,R_d$, that is, $S := R_1 \times \cdots \times R_d$.
We write $\operatorname{pr}_j : S \rightarrow R_j$ for the projection that extracts the
requirement $r_j$ on attribute $j$, so that $\operatorname{pr}_j(s) = r_j$ for every
$s = (r_1,\ldots,r_d) \in S$.

\subsubsection{Provisioning Cost}
\label{subsubsec:slice_provisioning}

\paragraph{Resources}

We assume that the 5G network consists of $p \in \mathbb{N}$ resource types,
such as the bandwidth of a link. 
For each resource $l \in [p]$, let $u_l \in \mathbb{R}_{\ge 0}$ be the largest
amount that may be allocated to a single slice.
We represent a resource allocation by
$\xi \!:=\! (\xi_1,\ldots,\xi_p) \!\in\! \mathbb{R}^{p}_{\ge 0}$, where $\xi_l$ is the
amount of resource $l \in [p]$ that is consumed and $\xi_l \le u_l$.
We then define $\Omega(s) \subseteq \mathbb{R}^{p}_{\ge 0}$ as the set
of such resource allocations that meet the requirements stated in the template
$s$.

\paragraph{Provisioning cost}

Let $\pi \in \mathbb{R}^{p}_{\ge 0}$ be the vector collecting the unit prices of
the $p$ resource types.
We define the resource allocation function
$\gamma : S \rightarrow \mathbb{R}^{p}_{\ge 0}$ by
\begin{align}
    \gamma(s) := \arg \min_{\xi \in \Omega(s)} \pi^{\top} \xi .
    \label{eq:gamma}
\end{align}
If multiple allocations attain the minimum, any of them may be chosen.
We make one further assumption on $\gamma$.

\begin{assumption}
\label{asm:gamma_blackbox}
No explicit form of $\gamma$ is available.
\end{assumption}

\noindent
Assumption~\ref{asm:gamma_blackbox} reflects the standardized
hierarchical architecture for slice management~\cite{etsi_zsm_003},
in which the slice operator is not provided with details of the
resource allocation algorithm.

Finally, we define the function $g : S \rightarrow \mathbb{R}_{\ge 0}$ that
returns the provisioning cost attributable to resource allocation:
\begin{align}
    g(s) := \pi^{\top} \gamma(s) .
    \label{eq:g_def}
\end{align}

\subsection{Dominance and Catalog-Based Request Handling}
\label{subsec:dominance_and_request_handling}

The selected catalog template must satisfy all
requested requirements. We first introduce template dominance to formalize
this condition (Sec.~\ref{subsubsec:dominance}), then define the catalog
and the cost of handling requests using the catalog
(Sec.~\ref{subsubsec:network_slice_catalog}).

\subsubsection{Dominance Relation}
\label{subsubsec:dominance}

For each attribute $j\in[d]$ and $r_1,r_2\in R_j$, let
$r_1\preceq_j r_2$ mean that every slice satisfying $r_2$ also satisfies
$r_1$. For example, a slice meeting a $5\,\mathrm{ms}$ latency requirement also
meets a $10\,\mathrm{ms}$ requirement, so
$10\,\mathrm{ms}\preceq_j 5\,\mathrm{ms}$. 
Under the following assumption on the attribute-wise relations
$\preceq_j$, we introduce dominance between templates.

\begin{assumption}
\label{asm:requirement_total_order}
For $j\in[d]$, $(R_j,\preceq_j)$ is a totally ordered set.
\end{assumption}

\begin{definition}
\label{def:dominance}
For $s_1,s_2\in S$, template $s_2$ \emph{dominates} $s_1$, written
$s_1\preceq s_2$, if
$\operatorname{pr}_j(s_1)\preceq_j \operatorname{pr}_j(s_2)$ for every $j\in[d]$.
\end{definition}

Thus, whenever $s_1\preceq s_2$, every resource allocation feasible for $s_2$
is also feasible for $s_1$, and hence
$\Omega(s_2)\subseteq\Omega(s_1)$. Since
$g(s)=\min_{\xi\in\Omega(s)}\pi^\top\xi$, this inclusion implies that
$g$ is monotone with respect to $\preceq$: $s_1\preceq s_2$ implies
$g(s_1)\leq g(s_2)$.

Although Assumption~\ref{asm:requirement_total_order} requires totally ordered
attributes, we can represent partially ordered requirements as a combination of totally ordered attributes.%
\footnote{Coverage requirements, for example, are naturally ordered
by set inclusion. For three base stations, represent a coverage requirement by
a binary vector with one coordinate per base station, where $1$ indicates that
coverage is required; thus, $(1,0,1)$ requires coverage at the first and third
stations. Treating these coordinates as separate attributes, each ordered by
$0\preceq 1$, satisfies Assumption~\ref{asm:requirement_total_order}, while the
induced componentwise order on the vectors coincides with set inclusion and
hence preserves the original partial order.}

\subsubsection{Network Slice Catalog and Request Handling}
\label{subsubsec:network_slice_catalog}
\label{subsubsec:formulation}

The slice operator fixes a catalog $U\subseteq S$ as a finite
subset of the template set $S$ of size $k\in\mathbb N$, so that $|U|=k$.
For a user-specified template $s\in S$, let
$U_s:=\{v\in U:s\preceq v\}$ be the catalog templates that satisfy its
requirements. 
Whereas $g(v)$ accounts for the resources allocated under template $v$,
$f:2^S\times S\rightarrow\mathbb R_{\ge0}$ gives the total provisioning cost of
request $s$ under catalog $U$:
\begin{align}
\label{eq:def_f}
f(U,s):=
\begin{cases}
\min_{v\in U_s}g(v),&U_s\neq\emptyset,\\
M,&U_s=\emptyset.
\end{cases}
\end{align}
Here $M \in \mathbb{R}$ is a fixed penalty approximating the total
provisioning cost of handling a request outside the catalog. It includes both
the resource cost and the additional operational cost of bespoke design and
validation passed on to the user.
We assume that $M\geq g(s)$ for $s\in S$.\footnote{The condition
$M\geq g(s)$ can be guaranteed without evaluating $g$: since $\pi\geq0$
and every feasible allocation is componentwise bounded by
$u=(u_1,\ldots,u_p)$, any $M\geq\pi^\top u$ satisfies this condition.}

\subsection{The \texorpdfstring{\acs{knscp}}{k-NSCP}}
\label{subsec:knscp}

\subsubsection{Formulation}
\label{subsubsec:knscp}
We assume that user-specified templates $s$ are \ac{iid} samples from a
distribution $\mathcal D$ over $S$. With the provisioning cost $f$
defined in~\eqref{eq:def_f}, the \ac{knscp} selects $k$ templates in advance
to minimize the expected provisioning cost under $\mathcal D$:
\begin{align}
\label{eq:knscp}
(\text{\acs{knscp}}):\quad
\min_{U\subseteq S,\,|U|=k}\mathbb E_{s\sim\mathcal D}[f(U,s)].
\end{align}

The formulation incorporates both catalog-design requirements.
Under~\eqref{eq:def_f}, a request $s$ served from the catalog is
assigned to a template in $U_s$ that satisfies all requirements of $s$,
ensuring \reqsla\ \textit{\ac{sla} satisfaction}.
The objective in~\eqref{eq:knscp} minimizes expected cost based on the
actual provisioning costs $g(v)$, thereby incorporating
\reqcost\ \textit{Provisioning-cost objective}.

\subsubsection{Sample Average Approximation}
\label{subsubsec:saa}
We approximate the \ac{knscp}~\eqref{eq:knscp} using \ac{saa},
a standard approach to stochastic programming. The \ac{saa}
assumes that a set $S_n:=\{s_1,\ldots,s_n\}\subseteq S$ of $n$
\emph{sample templates} distributed according to $\mathcal D$ is available.
It replaces $\mathbb E_{s\sim\mathcal D}[f(U,s)]$ in~\eqref{eq:knscp} by
the sample average:
\begin{align}
\label{eq:sampled_knscp}
(\text{\acs{saa}-\acs{knscp}}):\quad
\min_{U\subseteq S,\,|U|=k}\frac1n\sum_{i=1}^n f(U,s_i).
\end{align}
For notational convenience, let
$f_n(U):=n^{-1}\sum_{i=1}^n f(U,s_i)$ and let $f_n^\star$ denote the
optimal value of~\eqref{eq:sampled_knscp}.
When the standard assumptions of the \ac{saa}~\cite{Kim2024SAA} hold, the
sample optimum $f_n^\star$ converges in probability to the optimal value
of~\eqref{eq:knscp}.

The operator can obtain the sample templates from its request records,
because each user submits a user-specified template when requesting a slice
(Step~1 in Fig.~\ref{fig:catalog_design_policy}), including requests later
served outside the catalog.
For a new service without such records, representative requirements, for
example collected from prospective users, can be used instead.

\begin{figure}[t]
  \centering
  \includegraphics[width=0.8\columnwidth]{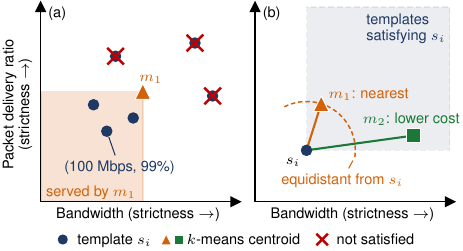}
\caption{Shortcomings of the $k$-means surrogate formulation for the
\acs{knscp}. (a) The centroid $m_1$ of the sample templates does not serve the
templates marked with $\times$, violating \reqsla\ \textit{\ac{sla}
satisfaction}. (b) For request $s$, nearest-centroid assignment may select $m_1$
rather than $m_2$, even when both satisfy $s$ and $g(m_1)>g(m_2)$, violating
\reqcost\ \textit{Provisioning-cost objective}.}
  \label{fig:distance_based_shortcomings}
\end{figure}

\begin{remark}
\label{rem:distance_based_formulations}
Distance-based clustering, including $k$-means~\cite{Lloyd1982,min2024}
and the metric $k$-median problem~\cite{hakimi1965median}, provides surrogate
approaches to the \ac{knscp}. They choose a catalog $U$ by
minimizing the distance from each sample to its nearest catalog template,
replacing the provisioning cost $f$ in~\eqref{eq:sampled_knscp} with
$\widetilde f_\rho(U,s):=\min_{v\in U}\rho(s,v)$.
For $k$-means,
$\rho(s,v)=\lVert\mathcal{E}(s)-\mathcal{E}(v)\rVert_2^2$, where
$\mathcal{E}:S\rightarrow\mathbb{R}^{d^{\prime}}$ embeds templates into a
Euclidean space; for the metric $k$-median problem, $\rho$ is a prescribed
metric on templates. These surrogates have two limitations.
For \reqsla\ \textit{\ac{sla} satisfaction}, proximity does not imply
dominance, so a selected template may fail to satisfy the request.
For \reqcost\ \textit{Provisioning-cost objective}, minimizing distance
does not minimize actual provisioning cost, even when the selected template
satisfies the request.
Figure~\ref{fig:distance_based_shortcomings}
illustrates both limitations using $k$-means.

\end{remark}

\standalonereferences

\section{Proposed Solution Method}
\label{sec:proposed_method}

We develop an iterative exact algorithm for the
\ac{saa}-\ac{knscp}~\eqref{eq:sampled_knscp}.
We first reformulate~\eqref{eq:sampled_knscp} and introduce the necessary
variables to facilitate the solution method (Sec.~\ref{subsec:assignment_cost_models}).
We then present the solution method (Sec.~\ref{subsec:solution_method})
and establish finite termination
(Sec.~\ref{subsec:guarantees}).
All proofs are given in Appendix~\ref{appendix:proposed_method_details}.

\subsection{Assignment Formulation and Cost Models}
\label{subsec:assignment_cost_models}

\subsubsection{Assignment Formulation}
\label{subsubsec:assignment_formulation}
Using the cost definition~\eqref{eq:def_f}, we express
the \ac{saa} problem~\eqref{eq:sampled_knscp}
in terms of template and sample-assignment decisions.
\textit{(i) Catalog templates:} For each $\ell\in[k]$, let $m_\ell\in S$
denote the $\ell$-th template in the catalog, and write
$m=(m_1,\ldots,m_k)$ for the resulting tuple.
\textit{(ii) Sample assignment:} Let $x_{i\ell}\in\{0,1\}$ indicate
that sample $s_i$ is assigned to $m_\ell$, and let $z_i\in\{0,1\}$
indicate that $s_i$ is handled outside the catalog and incurs the
penalty $M$. 
\begin{subequations}
\label{form:direct_attribute}
\begin{alignat}{2}
f_n^\star={}&
\mathrlap{\min_{m\in S^k,\,x,\,z}\
\frac1n\Bigl(\sum_{i,\ell}g(m_\ell)x_{i\ell}+M\sum_i z_i\Bigr)}
&&\label{obj:direct_attribute}\\
&\mathmakebox[\widthof{$\displaystyle\min_{m\in S^k}$}][c]{\text{s.t.}}\quad
&x_{i\ell}=1\Longrightarrow s_i\preceq m_\ell,
&\quad i\in[n],\ \ell\in[k]
\label{cons:direct_dominance}\\
&&\sum_{\ell=1}^k x_{i\ell}+z_i=1,
&\quad i\in[n]
\label{cons:sample_assignment}\\
&&x_{i\ell}\in\{0,1\},
&\quad i\in[n],\ \ell\in[k]
\label{cons:binary_x}\\
&&z_i\in\{0,1\},
&\quad i\in[n].
\label{cons:binary_z}
\end{alignat}
\end{subequations}

For a fixed catalog tuple $m\in S^k$, minimizing over $x$ and $z$
assigns each sample $s_i$ to a least-cost dominating template $m_\ell$, if one
exists, and
otherwise to the outside option. By~\eqref{eq:def_f} and
$g(m_\ell)\leq M$, its optimal value is therefore
$n^{-1}\sum_{i=1}^{n} f(\{m_1,\ldots,m_k\},s_i)
=f_n(\{m_1,\ldots,m_k\})$.
Although~\eqref{form:direct_attribute} optimizes over tuples
$m\in S^k$ that may contain repeated entries, it has the same
optimal value as~\eqref{eq:sampled_knscp}, which selects a
size-$k$ catalog $U\subseteq S$.\footnote{Every size-$k$ catalog
can be represented by a tuple in $S^k$. Conversely, given a tuple
$m\in S^k$, first remove repeated entries to form
$U_0:=\{m_1,\ldots,m_k\}$. Then obtain a size-$k$ catalog $U$ by
adding $k-|U_0|$ arbitrary distinct templates from $S\setminus U_0$.
By~\eqref{eq:def_f} and $g(v)\leq M$, adding templates cannot
increase the cost of any request, so $f_n(U)\leq f_n(U_0)$.}

\subsubsection{Cost Models}
\label{subsubsec:cost_models}

Let $y:=(x,z)$, and let $\mathcal Y$ be the finite set of binary assignments
satisfying~\eqref{cons:sample_assignment}. Our algorithm solves
the problem~\eqref{form:direct_attribute} under provisioning costs other than $g$, namely lower estimates. 
To describe both the true provisioning cost $g$ and its lower
estimates in a common form, we introduce a \emph{cost model}:
a monotone function $h:S\to[0,M]$ that assigns cost $h(v)$
to template $v$.
Under this model, we define the \emph{catalog cost} $\Phi_h(m,y)$, the
\emph{fixed-assignment cost} $\phi_h(y)$, and the \emph{optimal catalog cost} $\phi_h^\star$
by
\begin{align}
\Phi_h(m,y)&:=\frac1n\Bigl(\sum_{i,\ell}h(m_\ell)x_{i\ell}+M\sum_i z_i\Bigr),
\label{eq:catalog_cost}\\
\phi_h(y)&:=\min_{m\in S^k}\bigl\{\Phi_h(m,y):
\text{\eqref{cons:direct_dominance}}\bigr\},
\label{eq:fixed_assignment_cost}\\
\phi_h^\star&:=\min_{y\in\mathcal Y}\phi_h(y).
\label{eq:optimal_catalog_cost}
\end{align}
For the true cost model $g$, \eqref{form:direct_attribute} gives
$f_n^\star=\phi_g^\star$.

\subsection{Solution Method}
\label{subsec:solution_method}

\textbf{Challenges:}
Solving~\eqref{form:direct_attribute} remains challenging for two reasons.
(a) \textit{Black-box costs:} $g$ is available only through black-box
queries (Assumption~\ref{asm:gamma_blackbox}).
(b) \textit{Possibly infinite template space:} The template set $S$ may be
infinite.

\begin{algorithm}[t]
\caption{Catalog design with monotone lower bounds}
\label{algo:method}
\small
\begin{algorithmic}[1]
\Require Samples $S_n$, size $k$, tolerance $\varepsilon\geq0$,
  initial evaluated set $E_0$
\Ensure A catalog $U$ with a certified gap of at most $\varepsilon$
\State $E\gets E_0$, $\mathrm{UB}\gets+\infty$
\Repeat
  \State \(\triangleright\) \texttt{Master problem}
  \State Compute $\mathrm{LB}\gets \phi_{\underline g_E}^\star$ and an optimal assignment $\bar y$\label{line:method_master}
  \State \(\triangleright\) \texttt{Subproblem}
  \State Compute $\phi_g(\bar y)$ by evaluating $g$ on
    $J(\bar y)\setminus E$; $E\gets E\cup J(\bar y)$\label{line:method_subproblem}
  \If{$\phi_g(\bar y)<\mathrm{UB}$}
    \State $\mathrm{UB}\gets \phi_g(\bar y)$, $y^{\mathrm{inc}}\gets\bar y$\label{line:method_incumbent_update}
  \EndIf
\Until{$\mathrm{UB}-\mathrm{LB}\leq\varepsilon$}
\State Complete $J(y^{\mathrm{inc}})$ to a size-$k$ catalog $U$
\end{algorithmic}
\end{algorithm}

\textbf{Overview:}
Algorithm~\ref{algo:method} constructs a catalog $U$ whose
\emph{optimality gap}, $f_n(U)-f_n^\star$, is at most $\varepsilon\geq0$.
It keeps the set
$E\subseteq S$ of templates at which $g$ has been evaluated, and alternates
between a \emph{master problem} and a
\emph{subproblem}.

The master problem $\phi_{\underline g_E}^\star$ is obtained from $\phi_g^\star$
by replacing the black-box cost $g$ with an explicit optimistic
cost model $\underline g_E$, constructed from the evaluated costs
on $E$ and the monotonicity of $g$.
For the assignment $\bar y$ obtained from the master problem,
the subproblem computes the fixed-assignment cost $\phi_g(\bar y)$
under the true cost $g$.
The newly evaluated templates are added to $E$, and their costs
are used to update the optimistic cost model $\underline g_E$.
Table~\ref{tab:master_subproblem} summarizes the bounds,
solution methods, and required $g$ evaluations for the two problems.

\textbf{Why this method is effective:}
The algorithm evaluates $g$ only as needed and
uses the results to guide the search for a catalog.
It addresses (a) \textit{black-box costs} by accessing $g$ only through
oracle queries, without requiring an explicit functional form.
For (b) \textit{possibly infinite template space}, each catalog template
can be restricted to the upper template of the samples it serves, i.e., the
template taking the strictest value of each attribute among them
(Lemma~\ref{lem:join}), whose attribute values lie in the finite set of
observed values. Hence, the algorithm is
guaranteed to return a catalog with an optimality gap of at most
$\varepsilon$ after finitely many cost evaluations
(Theorem~\ref{thm:termination}).

\begin{table}[t]
\caption{Master problem and subproblem in each iteration.}
\label{tab:master_subproblem}
\centering
\footnotesize
\renewcommand{\arraystretch}{1.15}
\begin{tabular}{@{}P{0.18\linewidth}@{\hspace{5pt}}
  P{\dimexpr0.41\linewidth-5pt\relax}@{\hspace{5pt}}
  P{\dimexpr0.41\linewidth-5pt\relax}@{}}
\toprule
 & \textbf{Master problem:} $\phi_{\underline g_E}^\star$
 & \textbf{Subproblem:} $\phi_g(\bar y)$\\
\midrule
Bound on $f_n^\star$
  & Lower bound$^\dagger$: $\phi_{\underline g_E}^\star\leq f_n^\star$
  & Upper bound: $\phi_g(\bar y)\geq f_n^\star$\\
\addlinespace[2pt]
Solution method
  & \ac{milp}\newline
    (Sec.~\ref{subsubsec:milp}, Proposition~\ref{prop:milp})
  & Closed-form solution\newline
    (Sec.~\ref{subsubsec:join}, Lemma~\ref{lem:join})\\
\addlinespace[2pt]
$g$ evaluations
  & None
  & At most $k$\\
\bottomrule
\end{tabular}
\par\vspace{2pt}
\begin{minipage}{\linewidth}
\raggedright
$^\dagger$ $\phi_{\underline g_E}^\star$ is the largest lower bound on
$f_n^\star$ guaranteed by the evaluated costs on $E$ and
monotonicity alone
(Sec.~\ref{subsubsec:master}, Proposition~\ref{prop:best_bound}).
\end{minipage}
\end{table}

\subsubsection{Optimistic Cost}
\label{subsubsec:master}
We bound the unknown cost function $g$ from below using the optimistic
cost model $\underline g_E$.

\begin{definition}
\label{def:optimistic_cost}
The \emph{optimistic cost model} is
\begin{align}
\underline g_E(v):=\max\bigl(\{g(e):e\in E,\ e\preceq v\}\cup\{0\}\bigr),
\,\,v\in S.
\label{eq:optimistic_cost}
\end{align}
\end{definition}

By monotonicity, every template $v$ dominating some $e\in E$ satisfies
$g(v)\geq g(e)$; together with $g\geq0$, this gives $g\geq\underline g_E$.
We show that,
given only the evaluated costs and monotonicity, $\underline g_E$ is the
largest lower bound on $g$ (Lemma~\ref{lem:least_extension}) and the master
$\phi_{\underline g_E}^\star$ gives the largest lower bound on $f_n^\star$
(Proposition~\ref{prop:best_bound}).

\paragraph{Best lower bound on $g$}
Let $\mathcal H_E$ denote the set of monotone cost models that agree with
$g$ at every $e \in E$:
\begin{align}
\mathcal H_E\!:=\!\{h\!:\!S\!\to\![0,M]\text{ monotone}\mid h(e)\!=\!g(e),\ e\!\in\! E\}.
\label{eq:consistent_costs}
\end{align}
The following lemma establishes two properties of the optimistic
cost model $\underline g_E$ with respect to this class.

\begin{lemma}[Best lower bound on $g$]
\label{lem:least_extension}
The optimistic cost model $\underline g_E$ satisfies the following
two properties:
(i) \emph{Validity:}
$\underline g_E\leq h$ for every $h\in\mathcal H_E$.
(ii) \emph{Pointwise optimality:}
$\underline g_E$ is the pointwise largest such lower bound.
Moreover, $\underline g_E\in\mathcal H_E$.
\end{lemma}

\begin{figure}[t]
\centering
\includegraphics[width=0.73\columnwidth]{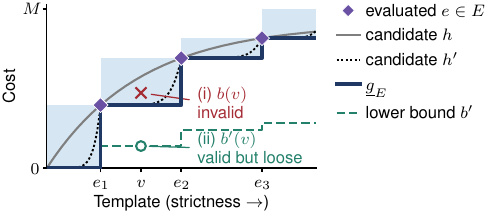}
\caption{Optimistic cost $\underline g_E$ for a one-dimensional $S$ with
$E=\{e_1,e_2,e_3\}$. Every candidate for the true cost, that is, every
$h\in\mathcal H_E$, passes through the shaded boxes; $h$ and $h'$ are two examples.
The marked comparisons illustrate (i) an invalid proposed lower bound $b$
and (ii) a loose lower bound $b'\leq\underline g_E$.}
\label{fig:optimistic_cost}
\end{figure}

Fig.~\ref{fig:optimistic_cost} illustrates the roles of these two
properties for a one-dimensional $S$.
\emph{(i) Validity} ensures that the lower bound holds regardless
of which $h\in\mathcal H_E$ is the true cost model.
The function $b$ fails this requirement: although $b(v)<h(v)$,
we have $b(v)>h'(v)$, so $b$ is not a lower bound if $g=h'$.
\emph{(ii) Pointwise optimality} ensures that the bound fully
exploits the evaluated costs and monotonicity.
The dashed $b'$ is a valid lower bound, but
$b'(v)<\underline g_E(v)$ shows that it can be tightened.

Lemma~\ref{lem:least_extension} thus identifies $\underline g_E$
as the largest lower bound on $g$ guaranteed by the evaluated costs
and monotonicity. We next use this result to show that the master
value $\phi_{\underline g_E}^\star$ is also the largest lower bound on $f_n^\star$
that can be guaranteed.

\paragraph{Best certified lower bound} 
We call $L\in\mathbb{R}$ a \emph{certified lower bound} for $E$
if $L\leq \phi_h^\star$ for every $h\in\mathcal H_E$; that is,
$L\leq f_n^\star$ regardless of which function in $\mathcal H_E$
is the true cost function.
The following proposition shows that $\phi_{\underline g_E}^\star$ is the best
certified lower bound for $E$, namely, the largest lower bound guaranteed
by the evaluated costs on $E$ and monotonicity alone.

\begin{proposition}
\label{prop:best_bound}
$\phi_{\underline g_E}^\star$ is a certified lower bound for $E$, and every certified
lower bound $L$ for $E$ satisfies $L\leq \phi_{\underline g_E}^\star = \mathrm{LB}$. 
\end{proposition}

Proposition~\ref{prop:best_bound} follows from
Lemma~\ref{lem:least_extension}, since $\phi_h^\star$ is monotone in $h$.
The gap in Algorithm~\ref{algo:method} decomposes as
$\mathrm{UB}-\mathrm{LB}=(\mathrm{UB}-\phi_g^\star)+(\phi_g^\star-\mathrm{LB})$.
From Proposition~\ref{prop:best_bound}, 
for the same evaluated set $E$, no other lower bound based only on
the evaluated costs and monotonicity can reduce the second term;
reducing it requires new evaluations or additional assumptions.

\subsubsection{Solution Method for the Master Problem}
\label{subsubsec:milp}
The master searches over the possibly infinite space $S^k$.
Exploiting the staircase structure of $\underline g_E$
(Fig.~\ref{fig:optimistic_cost}), we restrict each template to a finite
set of representatives without changing the optimum.

\begin{figure}[t]
\centering
\includegraphics[width=0.75\columnwidth]{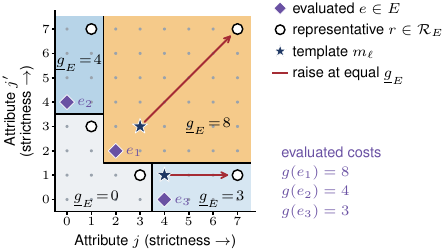}
\caption{Staircase of $\underline g_E$ and representatives for
$S=\{0,\ldots,7\}^2$ with every point as a sample and
$E=\{e_1,e_2,e_3\}$. Raising a template within a region where
$\underline g_E$ is constant leaves its cost unchanged and does not
reduce coverage.}
\label{fig:price_staircase}
\end{figure}

Fig.~\ref{fig:price_staircase} illustrates the idea behind this
reduction: \textit{within a region where $\underline g_E$ is constant, raising a template leaves its
cost unchanged and does not reduce sample coverage.}
To formalize this idea, let
$\mathrm{Cov}(v):=\{i\in[n]:s_i\preceq v\}$ be the samples that $v$ covers,
and let
$\mathcal P_E:=\{(\underline g_E(v),\mathrm{Cov}(v)):v\in S,\
\mathrm{Cov}(v)\neq\emptyset\}$
be the set of pairs of cost and covered samples. We call
$(c,K)\in\mathcal P_E$ \emph{maximal} if no $(c,K')\in\mathcal P_E$
satisfies $K\subsetneq K'$, that is, no template with the same cost covers a
strict superset of these samples.
We use these maximal pairs to define a finite set of representative
templates for the master problem.

\begin{definition}[Representatives]
\label{def:representatives}
A set $\mathcal R_E\subseteq S$ of \emph{representatives} contains exactly
one template $v$ with $(\underline g_E(v),\mathrm{Cov}(v))=(c,K)$ for each
maximal pair $(c,K)\in\mathcal P_E$.
\end{definition}

In Fig.~\ref{fig:price_staircase}, the representatives are $(1,3)$,
$(1,7)$, $(3,1)$, $(7,1)$, and $(7,7)$ (open circles).
Since $\underline g_E$ takes finitely many values,
$\mathcal R_E$ is finite. The master then satisfies the following.

\begin{proposition}[Representatives suffice]
\label{prop:milp}
The master problem has an optimal solution $(m,y)$ with
$m\in\mathcal R_E^k$.
\end{proposition}

By Proposition~\ref{prop:milp}, the search for the template tuple $m$
can be restricted from $S^k$ to $\mathcal R_E^k$ without changing
the master optimum. We solve the resulting finite selection and assignment
problem as an \ac{milp} (Appendix~\ref{appendix:master_milp}).
A concrete construction of $\mathcal R_E$ using the samples $S_n$
and the costs evaluated on $E$ is described in
Appendix~\ref{appendix:implementation_encoding}.

\subsubsection{Solution Method for the Subproblem}
\label{subsubsec:join}

For $y\in\mathcal Y$, let $n_\ell(y)$ denote the number of
samples assigned to template $m_\ell$, and let $C_\ell(y)$
denote the corresponding cluster:
\begin{align}
n_\ell(y)&\!:=\!\sum_{i=1}^n x_{i\ell},&\!\!
C_\ell(y)&\!:=\!\{s_i:i\in[n],\ x_{i\ell}=1\}.
\end{align}
For monotone $h$, the upper template defined below yields an optimal
template tuple for $\phi_h(y)$ in~\eqref{eq:fixed_assignment_cost} and a
closed-form expression for its value (Lemma~\ref{lem:join}).

\begin{definition}
\label{def:upper_template}
The \emph{upper template} of a nonempty finite set $T\subseteq S$,
denoted by $\bigvee T$, is formed by taking the maximum value of
each attribute with respect to its order $\preceq_j$:
\begin{align}
\label{eq:upper_template}
\bigvee T\!:=\!\left(\!
\max_{\preceq_1}\{\operatorname{pr}_1(s)\!:\!s\!\in\! T\},\ldots,
\max_{\preceq_d}\{\operatorname{pr}_d(s)\!:\!s\!\in\! T\}
\!\right).
\end{align}
\end{definition}

\begin{lemma}
\label{lem:join}
Let $h:S\to[0,M]$ be monotone. For every $y\in\mathcal Y$, an optimal
template tuple $m^\star$ for $\phi_h(y)$ is obtained by setting
$m^\star_\ell=\bigvee C_\ell(y)$ for every $\ell\in[k]$ with
$n_\ell(y)>0$. For every $\ell\in[k]$ with
$n_\ell(y)=0$, $m^\star_\ell$ may be chosen arbitrarily from $S$.
The optimal value is
\begin{align}
\label{eq:exact_assignment_value}
\phi_h(y)=\frac1n\left(
\sum_{\ell:n_\ell(y)>0}\!n_\ell(y)
h(\bigvee C_\ell(y))
\!+\!M\sum_{i=1}^n z_i\right).
\end{align}
\end{lemma}

\begin{figure}[t]
\centering
\includegraphics[width=0.62\columnwidth]{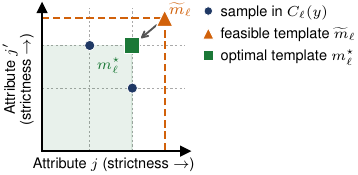}
\caption{Optimality of the upper template. Any feasible template
$\widetilde m_\ell$ dominates every template in $C_\ell(y)$ and therefore
satisfies $\bigvee C_\ell(y)\preceq\widetilde m_\ell$. Thus, for any
monotone cost $h$, $h(m^\star_\ell)\leq h(\widetilde m_\ell)$ for
$m^\star_\ell=\bigvee C_\ell(y)$.}
\label{fig:induced_upper_template}
\end{figure}

Figure~\ref{fig:induced_upper_template} illustrates the key idea of Lemma~\ref{lem:join}: 
the upper template $\bigvee C_\ell(y)$ of the cluster $C_\ell(y)$ dominates every template in $C_\ell(y)$, 
and has the least cost
among all templates that dominate every sample in $C_\ell(y)$.

For a fixed assignment $y$, the subproblem computes $\phi_g(y)$ by evaluating
$g$ only on the set
$J(y)\!:=\!\{\bigvee\! C_\ell(y)\!:\!\ell\!\in\![k],\ n_\ell(y)\!>\!0\}$.
This requires at most $k$ evaluations of $g$.

\subsection{Guarantees}
\label{subsec:guarantees}
Theorem~\ref{thm:termination} guarantees that finitely many $g$ evaluations
suffice to obtain a catalog with an optimality gap of at most $\varepsilon$.
Let $V_j:=\{\operatorname{pr}_j(s_i):i\in[n]\}$ be the observed values of attribute $j$, and
let $G_n:=V_1\times\cdots\times V_d$. 

\begin{theorem}
\label{thm:termination}
Suppose that $g$ is evaluated exactly and each master is solved exactly. Then
Algorithm~\ref{algo:method} terminates after at most $|G_n\setminus E_0|+1$
iterations and makes at most
$|G_n\setminus E_0|\leq\prod_{j}|V_j|\leq n^d$ new evaluations of $g$. The returned
catalog $U$ satisfies $f_n(U)-f_n^\star\leq\varepsilon$.
\end{theorem}

\standalonereferences

\providecommand{\CurrentMSTenThree}{0.29}
\providecommand{\CurrentMSTenFive}{14.39}
\providecommand{\CurrentSmallInstances}{20}
\providecommand{\CurrentSmallLtwoHits}{4}
\providecommand{\CurrentSmallLtwoError}{5.92}
\providecommand{\CurrentSmallHellingerHits}{0}
\providecommand{\CurrentSmallHellingerError}{31.47}
\providecommand{\CurrentRingMSThree}{0.28}
\providecommand{\CurrentRingMSFive}{14.54}
\providecommand{\RefinedAnyTrainStart}{2.164}
\providecommand{\RefinedAnyTrainEnd}{1.810}
\providecommand{\RefinedAnyTestStart}{2.204}
\providecommand{\RefinedAnyTestEnd}{1.862}
\providecommand{\RefinedAnyEndCalls}{185}
\providecommand{\RefinedAnyTimeLimited}{3}
\providecommand{\RefinedAnyReachedCap}{7}
\providecommand{\RefinedSmallMinCalls}{13}
\providecommand{\RefinedSmallMaxCalls}{35}
\providecommand{\RefinedMainMinGap}{26.94}
\providecommand{\RefinedMainMaxGap}{32.06}

\section{Experiments}
\label{sec:experiment}

\subsection{Experimental Setup}
\label{subsec:experiment_setup}

\subsubsection{Environment}
We use fixed tree and ring networks, each with one core, four transport
routers, and twelve \acp{gnb}, with shortest-path routing.
The traffic, queueing, and resource-allocation models
follow~\cite{kobayashi2026decomposition}. Each request uses all twelve
\acp{gnb}. We emulate user demand by an equal-probability mixture of three
service families representing remote operation, XR video, and buffered media,
from which both the sample templates $S_n$ and the test requests are drawn.
Appendix~\ref{appendix:experiment_details} gives the detailed parameters.

We approximate $g(s)$ by taking the minimum cost $\pi^\top\xi$ among feasible
allocations $\xi\in\Omega(s)$ obtained from 100 allocation trials in the
simulator. 

\subsubsection{Comparison Methods}
We compare the following five methods. H. abbreviates Hellinger and MS
denotes multi-start.

\noindent\textbf{(i) L2, (ii) Hellinger.}
Apply $k$-means to $S_n$ using squared L2 or Hellinger-based
distance~\cite{min2024}, respectively. Select the template in $S_n$ nearest
each centroid as its cluster's representative, and form the catalog $U$
as the set of these selected templates.

\noindent\textbf{(iii) L2+Join+MS, (iv) H.+Join+MS.}
These methods extend L2 and Hellinger, respectively, in two ways.
\textit{Join} replaces each cluster's representative with its upper
template~\eqref{eq:upper_template}, ensuring \ac{sla} satisfaction
for all requests in that cluster.
\textit{MS} repeats this catalog construction with different clustering
seeds and retains the catalog with the lowest training cost $f_n(U)$.

\noindent\textbf{(v) Ours.}
Apply Algorithm~\ref{algo:method} with $\varepsilon=0$, using the
initialization and simple enhancements described in
Appendix~\ref{appendix:experimental_implementation}.

\subsubsection{Protocol and Measures}
Each condition uses ten paired seeds: within a seed, all methods use the
same sample templates $S_n$ and, where test cost is reported, the same
$N=10{,}000$ test requests across methods and networks.
We set $\pi_l=1$ for all $l\in[p]$ and the penalty
$M=10^{10}$.

\noindent\textbf{(i) Test cost and (ii) Coverage.}
Test cost is the held-out mean
$N^{-1}\sum_{s\in S_{\mathrm{test}}}f(U,s)$. Cov. reports the percentage of test
requests served by the catalog. 

\noindent\textbf{(iii) Calls, (iv) Algorithm time, and (v) Gap.}
Calls denotes the number of actual evaluations of $g$, including initialization.
$T_{\rm alg}$ is elapsed catalog-design time minus time inside $g$ calls. 
Gap is the proposed algorithm's relative optimality gap
$(\mathrm{UB}-\mathrm{LB})/\mathrm{UB}$.

\begin{table*}[t]
\centering
\caption{Downlink workload on tree and ring networks.
Cost is reported as mean $\pm$ sample standard deviation;
others are means.}
\label{tab:experiment_downlink}
\footnotesize
\scalebox{0.9}{%
  \begin{tabular*}{\textwidth}{@{\extracolsep{\fill}}clrrrrrrrrrr@{}}
\toprule
 & & \multicolumn{5}{c}{Tree} & \multicolumn{5}{c}{Ring}\\
\cmidrule(lr){3-7}\cmidrule(l){8-12}
$k$ & Method & Cost ($\times 10^9$) & Cov. (\%) & Calls & $T_{\rm alg}$ (s) & Gap (\%) & Cost ($\times 10^9$) & Cov. (\%) & Calls & $T_{\rm alg}$ (s) & Gap (\%)\\
\midrule
\multirow{5}{*}{3} & L2 & 8.094 $\pm$ 0.396 & 22.34 & 3.0 & 0.08 & -- & 8.214 $\pm$ 0.385 & 22.34 & 3.0 & 0.09 & --\\
 & Hellinger & 8.600 $\pm$ 0.604 & 18.01 & 3.0 & 0.09 & -- & 8.737 $\pm$ 0.519 & 18.01 & 3.0 & 0.07 & --\\
 & L2+Join+MS & 2.252 $\pm$ 0.016 & 99.45 & 3.4 & 50.0 & -- & 3.059 $\pm$ 0.020 & 99.45 & 3.4 & 50.0 & --\\
 & H.+Join+MS & 2.958 $\pm$ 0.021 & 99.80 & 3.7 & 50.0 & -- & 4.063 $\pm$ 0.030 & 99.80 & 3.7 & 50.0 & --\\
 & Ours & \textbf{2.245} $\pm$ 0.019 & 99.45 & 49.0 & 1.00 & 30.56 & \textbf{3.050} $\pm$ 0.028 & 99.45 & 49.0 & 0.95 & 32.06\\
\midrule
\multirow{5}{*}{5} & L2 & 7.218 $\pm$ 0.254 & 34.15 & 5.0 & 0.09 & -- & 7.441 $\pm$ 0.247 & 34.15 & 5.0 & 0.07 & --\\
 & Hellinger & 7.316 $\pm$ 0.821 & 31.47 & 5.0 & 0.10 & -- & 7.489 $\pm$ 0.774 & 31.47 & 5.0 & 0.09 & --\\
 & L2+Join+MS & 2.187 $\pm$ 0.031 & 99.47 & 10.4 & 50.0 & -- & 2.973 $\pm$ 0.039 & 99.47 & 10.5 & 50.0 & --\\
 & H.+Join+MS & 2.483 $\pm$ 0.223 & 99.80 & 10.3 & 50.0 & -- & 3.428 $\pm$ 0.298 & 99.80 & 10.3 & 50.0 & --\\
 & Ours & \textbf{1.872} $\pm$ 0.043 & 99.28 & 50.0 & 0.69 & 26.94 & \textbf{2.541} $\pm$ 0.052 & 99.28 & 50.0 & 0.71 & 28.07\\
\bottomrule
\end{tabular*}
}
\end{table*}

\subsection{Catalog-Size and Topology Comparison}
\label{subsec:experiment_nk}

We assess held-out catalog quality and algorithm time across
methods as catalog size and network topology vary.
We use $n=1000$ and $k\in\{3,5\}$.
All methods are evaluated under common budgets of
50 seconds of algorithm time $T_{\rm alg}$ and
50 distinct evaluations of $g$, including initialization.

\subsubsection{Result}
Table~\ref{tab:experiment_downlink} shows that our method reduces
Test cost relative to L2+Join+MS by \CurrentMSTenFive\% on tree and
\CurrentRingMSFive\% on ring at $k=5$.
At $k=5$, our method achieves lower Test cost than L2+Join+MS
in all ten paired seeds on both networks.
Gains at $k=3$ are smaller: \CurrentMSTenThree\% on tree and \CurrentRingMSThree\% on ring.
The proposed method's mean relative gaps in Table~\ref{tab:experiment_downlink}
remain \RefinedMainMinGap--\RefinedMainMaxGap\%, so these large-instance results are feasible incumbents
without global-optimality certificates.

\subsubsection{Discussion}
Our method achieves larger Test-cost reductions over L2+Join+MS at $k=5$
than at $k=3$. A possible explanation lies in the workload's three-family
structure. At $k=3$, distance-based clustering may group requests with
similar requirements according to these service families.
Taking each cluster's upper template may then yield
\ac{sla}-satisfying representatives \reqsla with limited
overprovisioning, leaving little room for further cost reduction.
At $k=5$, however, the three-family structure alone no longer
identifies a natural partition into five clusters. This may explain
the larger benefit our method achieves over L2+Join+MS at $k=5$.

These observations suggest
that clustering may achieve a Test cost close to that of our method with
fewer cost evaluations when $k$ matches the number of service families.
When the number of demand groups differs from $k$ or is unknown,
our method may offer greater cost reductions.

\subsection{Anytime Improvement with Cost Evaluations}
\label{subsec:experiment_anytime}

We examine how our method reduces Train and Test costs
as the number of cost evaluations increases. We use the tree network
at $(n,k)=(1000,5)$, with budgets of 200 evaluations of $g$ and 50 algorithm
seconds excluding oracle calls.

\subsubsection{Result}
\begin{figure}[t]
\centering
\includegraphics[width=0.8\columnwidth]{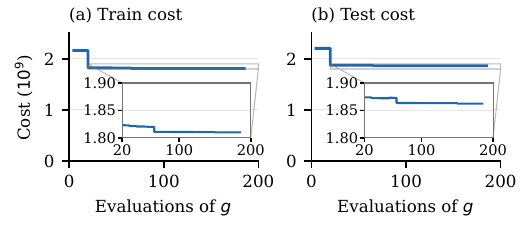}
\caption{Mean training and test costs against cumulative evaluations of $g$
on tree for the proposed method. 
All initialization calls count toward the budget.
Insets magnify the 20--200-evaluation range.
}
\label{fig:experiment_anytime}
\end{figure}

Fig.~\ref{fig:experiment_anytime} shows mean Train and Test costs up to
\RefinedAnyEndCalls{} evaluations, the largest count reached by all ten runs
within the time budget. From initialization to this point, mean Train cost decreases
from $\RefinedAnyTrainStart$ to $\RefinedAnyTrainEnd$, and Test cost
from $\RefinedAnyTestStart$ to $\RefinedAnyTestEnd$, in units of
$10^9$.
Most of the improvement occurs within the first 20 evaluations in both Train
and Test averages.

\subsection{Convergence on Small Instances}
\label{subsec:experiment_exactness}

We examine whether our method can find a globally optimal catalog in
practice by checking whether $\mathrm{UB}-\mathrm{LB}$ decreases to zero
on small instances.
We use the tree network with ten seeds and set $n=20$ and $k\in\{3,5\}$.

\begin{figure}[t]
\centering
\includegraphics[width=0.8\columnwidth]{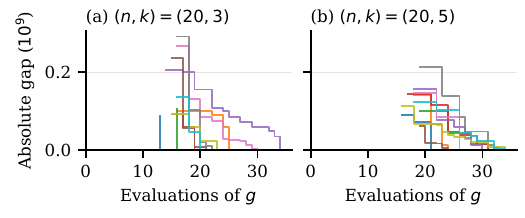}
\caption{Optimality gap $\mathrm{UB}-\mathrm{LB}$ at $n=20$.
Each panel contains ten seeds.}
\label{fig:experiment_exactness}
\end{figure}

\subsubsection{Result}
Fig.~\ref{fig:experiment_exactness} shows that the proposed method
reduces its absolute optimality gap to at most $10^{-6}$ cost units
in all \CurrentSmallInstances{} instances, using
\RefinedSmallMinCalls--\RefinedSmallMaxCalls{} cost evaluations.

\standalonereferences

\section{Conclusion}
\label{sec:conclusion}

We formulated the \ac{knscp} to minimize expected provisioning cost, and developed a solution method for its \ac{saa}
problem that replaces the black-box provisioning cost with an optimistic cost
model built from the evaluated costs and monotonicity. 
In experiments, our method reduced held-out cost by \CurrentMSTenFive\% on tree and
\CurrentRingMSFive\% on ring, and
reduced its optimality gap to zero on all \CurrentSmallInstances{}
small instances.
Closing large-instance optimality gaps and evaluating more than two varying
request coordinates remain future work.

\standalonereferences

\appendices

\section{Proofs for the Proposed Method}
\label{appendix:proposed_method_details}

\subsection{Proofs of Lemma~\ref{lem:least_extension} and
Proposition~\ref{prop:best_bound}}
\label{appendix:proof_best_bound}

\begin{proof}[Proof of Lemma~\ref{lem:least_extension}]
\emph{$\underline g_E\in\mathcal H_E$:}
We verify the three requirements in the definition of $\mathcal H_E$.
\begin{itemize}
\item \emph{Monotonicity.}
If $v\preceq v'$, every $e\in E$ with $e\preceq v$
also satisfies $e\preceq v'$. Hence
\[
\{e\in E:e\preceq v\}
\subseteq
\{e\in E:e\preceq v'\}.
\]
By the definition of $\underline g_E$
in~\eqref{eq:optimistic_cost}, this inclusion implies
$\underline g_E(v)\leq\underline g_E(v')$.

\item \emph{Agreement with evaluated costs.}
Fix $e\in E$. Since $e\preceq e$, the maximum defining
$\underline g_E(e)$ includes $g(e)$, so
$\underline g_E(e)\geq g(e)$.
Conversely, every $e'\in E$ with $e'\preceq e$ satisfies
$g(e')\leq g(e)$ by monotonicity of $g$.
Together with $0\leq g(e)$, this gives
$\underline g_E(e)\leq g(e)$.
Hence $\underline g_E(e)=g(e)$.

\item \emph{Range.}
Every value in the maximum defining $\underline g_E(v)$
belongs to $[0,M]$, including the always-present value $0$.
Therefore $0\leq\underline g_E(v)\leq M$ for every $v\in S$.
\end{itemize}

\emph{(i) Validity:}
For any $h\in\mathcal H_E$ and
$e\in E$ with $e\preceq v$, $h(v)\geq h(e)=g(e)$; together with $h\geq0$,
this gives $h(v)\geq\underline g_E(v)$. Thus $\underline g_E$ bounds every
$h\in\mathcal H_E$ from below.

\emph{(ii) Pointwise optimality:}
If $\psi\leq h$ for every $h\in\mathcal H_E$,
choosing $h=\underline g_E\in\mathcal H_E$ gives $\psi\leq\underline g_E$.
Since $\underline g_E$ is itself a common lower bound, it is therefore
the largest common lower bound of $\mathcal H_E$.
\end{proof}

\begin{proof}[Proof of Proposition~\ref{prop:best_bound}]
Since $x,z\geq0$, $h\leq h'$ pointwise implies
$\Phi_h(m,y)\leq\Phi_{h'}(m,y)$ for all $(m,y)$, and hence
$\phi_h\leq \phi_{h'}$ and $\phi_h^\star\leq \phi_{h'}^\star$. For every $h\in\mathcal H_E$,
$\underline g_E\leq h$ by Lemma~\ref{lem:least_extension}, so $\phi_{\underline g_E}^\star\leq \phi_h^\star$. Thus
$\phi_{\underline g_E}^\star$ is a certified lower bound. Since $g\in\mathcal H_E$, in
particular $\phi_{\underline g_E}^\star\leq \phi_g^\star=f_n^\star$. For any certified lower bound
$L$, choosing $h=\underline g_E\in\mathcal H_E$ gives
$L\leq \phi_{\underline g_E}^\star$.
\end{proof}

\subsection{Proof of Proposition~\ref{prop:milp}}
\label{appendix:proof_milp}

\begin{proof}[Proof of Proposition~\ref{prop:milp}]
\emph{(i) Existence of an optimal solution:}
By Lemma~\ref{lem:least_extension}, $\underline g_E$ is a monotone cost
model. Hence Lemma~\ref{lem:join} guarantees that, for every
$y\in\mathcal Y$, the minimum over $m$ defining
$\phi_{\underline g_E}(y)$ is attained. Since $\mathcal Y$ is finite, some
$y\in\mathcal Y$ satisfies $\phi_{\underline g_E}(y)=\phi_{\underline g_E}^\star$.
Choosing a corresponding minimizing template tuple $m$ gives an optimal
solution $(m,y)$.

\emph{(ii) Construction of an optimal solution using representatives:}
Starting from the optimal solution $(m,y)$ obtained in (i),
we replace each template $m_\ell$ by a representative in $\mathcal R_E$
while preserving feasibility and the objective value.
Fix $\ell\in[k]$. We consider two cases.
\begin{itemize}
\item \emph{Case $n_\ell(y)>0$:}
Template $m_\ell$ covers its assigned samples, so $\mathrm{Cov}(m_\ell)\neq\emptyset$
and $(c,\mathrm{Cov}(m_\ell))\in\mathcal P_E$ with
$c:=\underline g_E(m_\ell)$. Since $\mathcal P_E$ is finite, some maximal
pair $(c,K)$ satisfies $\mathrm{Cov}(m_\ell)\subseteq K$; let
$r\in\mathcal R_E$ be its representative
(Definition~\ref{def:representatives}). Replacing $m_\ell$ by $r$ keeps
every assigned sample covered and leaves
$\Phi_{\underline g_E}(m,y)$ unchanged.

\item \emph{Case $n_\ell(y)=0$:}
Template $m_\ell$ affects neither~\eqref{cons:direct_dominance} nor
$\Phi_{\underline g_E}(m,y)$, so it can be set to any representative;
$\mathcal R_E\neq\emptyset$ because
$(\underline g_E(s_1),\mathrm{Cov}(s_1))\in\mathcal P_E$.
\end{itemize}
Applying these replacements for every $\ell\in[k]$ preserves feasibility
and the objective value, yielding an optimal solution with
$m\in\mathcal R_E^k$.
\end{proof}

\subsection{Proof of Lemma~\ref{lem:join}}
\label{appendix:proof_join}

\begin{proof}
Fix $y\in\mathcal Y$. Templates $m_\ell$ with $n_\ell(y)=0$ affect neither
constraint~\eqref{cons:direct_dominance} nor $\Phi_h(m,y)$. For any $\ell$ with $n_\ell(y)>0$,
constraint~\eqref{cons:direct_dominance} requires $m_\ell$ to dominate every
template in $C_\ell(y)$. Since each $(R_j,\preceq_j)$ is totally ordered
and~\eqref{eq:upper_template} takes the attribute-wise maximum, this holds
if and only if $\bigvee C_\ell(y)\preceq m_\ell$. Monotonicity of $h$ then
gives $h(m_\ell)\geq h(\bigvee C_\ell(y))$, with equality at
$m_\ell=\bigvee C_\ell(y)$. Substituting these templates into $\Phi_h$ gives
the stated value, so the minimum defining $\phi_h(y)$ is attained. 
\end{proof}

\subsection{Proof of Theorem~\ref{thm:termination}}
\label{appendix:proof_termination}

\begin{proof}
\emph{(i) Termination and evaluation count:} Consider an iteration with 
assignment $\bar y$. Since $\bar y$ is optimal for the master,
$\phi_{\underline g_E}(\bar y)=\mathrm{LB}$, so Lemma~\ref{lem:join} applied to
$g$ and to $\underline g_E$ gives
\begin{align}
\phi_g(\bar y)-\mathrm{LB}=\frac1n\sum_{\ell:n_\ell(\bar y)>0}n_\ell(\bar y)
\bigl(g-\underline g_E\bigr)\bigl(\textstyle\bigvee C_\ell(\bar y)\bigr)\geq0.
\notag
\end{align}
If $g=\underline g_E$ on $J(\bar y)$, then $\phi_g(\bar y)=\mathrm{LB}$, so
$\mathrm{UB}\leq\mathrm{LB}$ after the update and the algorithm stops. Hence every iteration that does not stop finds an upper
template with $g>\underline g_E$. Since $\underline g_E=g$ on $E$, this
template is not yet evaluated, and it lies in $G_n$. Each iteration that does
not stop therefore adds at least one new point of $G_n$ to $E$.
There are at most $|G_n\setminus E_0|$ such iterations, so the algorithm
terminates after at most $|G_n\setminus E_0|+1$ iterations, including
the final iteration in which it stops. Since all newly evaluated templates
lie in $G_n\setminus E_0$ and no template is evaluated more than once,
the number of new evaluations of $g$ is at most
$|G_n\setminus E_0|\leq\prod_j|V_j|\leq n^d$.

\emph{(ii) Certified gap:} Since $\mathrm{UB}=+\infty$ initially, the first
iteration sets $y^{\mathrm{inc}}$, so it is defined at termination. At
termination, $\mathrm{LB}\leq f_n^\star$ by
Proposition~\ref{prop:best_bound}, $f_n^\star\leq f_n(U)$ because $U$ is a
feasible catalog, and $f_n(U)\leq\mathrm{UB}$ because $y^{\mathrm{inc}}$ is
feasible under $U$ and reassigning samples cannot increase the cost. With
$\mathrm{UB}\leq\mathrm{LB}+\varepsilon$, we obtain
$f_n(U)-f_n^\star\leq\varepsilon$.
\end{proof}

\standalonereferences

\section{Master Problem Formulation and Representative Construction}
\label{appendix:implementation_refinements}

We formulate the master $\phi_{\underline g_E}^\star$ as an
\ac{milp} (Appendix~\ref{appendix:master_milp}) and describe how to construct the representatives $\mathcal R_E$
from observed attribute values when $E\subseteq G_n$ (Appendix~\ref{appendix:implementation_encoding}).

\subsection{Master MILP Formulation}
\label{appendix:master_milp}

The master problem $\phi_{\underline g_E}^\star$ chooses a catalog tuple
$m=(m_1,\ldots,m_k)\in S^k$ and determines whether each sample $s_i$
is assigned to a catalog template $m_\ell$ ($x_{i\ell}=1$)
or handled outside the catalog ($z_i=1$).
By Proposition~\ref{prop:milp}, the search for $m$ can be restricted
from $S^k$ to $\mathcal R_E^k$ without changing the optimum.
With this finite candidate set, we represent catalog selection
by $a_r\in\{0,1\}$, indicating whether representative $r\in\mathcal R_E$
is selected, and sample assignment by $x_{ir}\in\{0,1\}$, defined only
for pairs $s_i\preceq r$, indicating assignment of $s_i$ to $r$, while
retaining $z_i$. The resulting formulation is
\begin{subequations}
\label{form:master_milp}
\begin{alignat}{2}
&\mathrlap{\phi_{\underline g_E}^\star=\min_{a,x,z}\quad\frac1n\Bigl(
\sum_i\sum_{\substack{r\in\mathcal R_E\\s_i\preceq r}}
\underline g_E(r)\,x_{ir}+M\sum_i z_i\Bigr),}&&\\
&\text{s.t.}\quad\sum_{r\in\mathcal R_E}a_r\leq k,&&
\label{cons:candidate_cardinality}\\
&\phantom{\text{s.t.}\quad}
\sum_{\substack{r\in\mathcal R_E\\s_i\preceq r}}x_{ir}+z_i=1,
&\quad&i\in[n],
\label{cons:candidate_assignment}\\
&\phantom{\text{s.t.}\quad}
x_{ir}\leq a_r,&\quad&i\in[n],\ r\in\mathcal R_E,\ s_i\preceq r,
\label{cons:candidate_activation}\\
&\phantom{\text{s.t.}\quad}
x_{ir}\in\{0,1\},&\quad&i\in[n],\ r\in\mathcal R_E,\ s_i\preceq r,\\
&\phantom{\text{s.t.}\quad}
a_r\in\{0,1\},&\quad&r\in\mathcal R_E,\\
&\phantom{\text{s.t.}\quad}
z_i\in\{0,1\},&\quad&i\in[n].
\end{alignat}
\end{subequations}
Constraint~\eqref{cons:candidate_cardinality} limits the number of
selected representatives to $k$, since a tuple $m\in\mathcal R_E^k$
contains at most $k$ distinct representatives.
Each $a_r$ counts representative $r$ only once, even if $r$ occurs
repeatedly in $m$: all samples assigned to those occurrences can be
assigned to a single copy without changing feasibility or cost.
Constraint~\eqref{cons:candidate_assignment} is the counterpart of
\eqref{cons:sample_assignment}, requiring each sample to be assigned
either to one selected representative or to the outside option.
Constraint~\eqref{cons:candidate_activation} permits assignment
only to a selected representative; coverage is enforced by
defining $x_{ir}$ only when $s_i\preceq r$.

Take an optimal solution of~\eqref{form:master_milp} and place its
selected representatives in distinct positions of a tuple
$m\in\mathcal R_E^k$, filling the remaining positions with arbitrary
representatives to which no samples are assigned.
Relabeling the assignments by these positions yields an optimal master
solution $(m,\bar y)$ with $\bar y\in\mathcal Y$.

\begin{remark}[Continuous assignment variables]
The integrality constraints on $x,z$ in~\eqref{form:master_milp} can be
relaxed to $0\leq x_{ir}\leq1$ and $0\leq z_i\leq1$ without changing
the optimal value. To see this, separate the selection of $a$ from the
optimization over $x,z$:
\begin{multline*}
\phi_{\underline g_E}^\star
\!=\!\min_{\substack{a_r\in\{0,1\}\\
\text{\eqref{cons:candidate_cardinality}}}}
\;\min_{\substack{x_{ir},z_i\in\{0,1\}\\
\text{\eqref{cons:candidate_assignment}, \eqref{cons:candidate_activation}}}}
\frac1n\sum_i
\underbrace{\Bigl(
\sum_{\substack{r\in\mathcal R_E\\s_i\preceq r}}
\underline g_E(r)x_{ir}\!+\!Mz_i\Bigr)}_{\text{(a)}}.
\end{multline*}
For each fixed feasible $a$, the relaxed assignment problem
separates by sample.
For each sample $i$, term~(a) is its assignment cost.
The assignment weights $x_{ir}$ and $z_i$ sum to one
by~\eqref{cons:candidate_assignment}, so term~(a) is a weighted average
of the costs $\underline g_E(r)$ of the selected representatives that
cover $s_i$ and the outside-option cost $M$.
This average cannot be smaller than the
cheapest of these costs, which is attained by assigning the sample
entirely to a cheapest option. Hence, for each fixed feasible $a$,
an optimal binary assignment exists even when the binary requirements
are replaced by $0\leq x_{ir}\leq1$ and $0\leq z_i\leq1$.
The inner minimum, and therefore the outer minimum
$\phi_{\underline g_E}^\star$, is unchanged. 
\end{remark}

\subsection{Construction of Representatives}
\label{appendix:implementation_encoding}
We construct a finite representative set $\mathcal R_E$ satisfying
Definition~\ref{def:representatives} from the samples $S_n$ and the costs already
evaluated on $E$. The construction has three steps, and each step is justified below.
Throughout, let $V_j:=\{\operatorname{pr}_j(s_i):i\in[n]\}$ and
$G_n:=V_1\times\cdots\times V_d$, as in Sec.~\ref{subsec:guarantees},
and suppose $E\subseteq G_n$.
Recall that $\mathrm{Cov}(v):=\{i\in[n]:s_i\preceq v\}$.

\paragraph{Step 1: Restrict to observed values}
\begin{lemma}[Restriction to observed values]
\label{lem:observed_value_restriction}
For every $v\in S$ with $\mathrm{Cov}(v)\neq\emptyset$, there exists
$\widehat v\in G_n$ such that
\[
\mathrm{Cov}(\widehat v)=\mathrm{Cov}(v),
\qquad \underline g_E(\widehat v)=\underline g_E(v).
\]
\end{lemma}
\begin{proof}
For each attribute $j$, round $\operatorname{pr}_j(v)$ down to the greatest observed
value no greater than it in the requirement order:
\[
\operatorname{pr}_j(\widehat v):=
\max_{\preceq_j}\{t\in V_j:t\preceq_j \operatorname{pr}_j(v)\}.
\]
The set is nonempty because $v$ covers a sample.
For every $u\in G_n$, each coordinate of $u$ is an observed value, so
$u\preceq v$ if and only if $u\preceq\widehat v$.
Applying this equivalence to the samples preserves $\mathrm{Cov}(v)$.
Applying it to $e\in E\subseteq G_n$ preserves the evaluated templates
in the maximum defining $\underline g_E$ in~\eqref{eq:optimistic_cost},
and hence preserves its value.
\end{proof}
Thus restricting the template space from $S$ to $G_n$ preserves every
pair of optimistic cost $\underline g_E(v)$ and covered sample
set $\mathrm{Cov}(v)$, without choosing a discretization resolution.

\paragraph{Step 2: Keep one candidate per cell}
For each attribute $j$, sort $V_j$ in the order $\preceq_j$ and split
it immediately before each distinct value $\operatorname{pr}_j(e)$ with $e\in E$.
Each boundary value belongs to the block beginning at that value;
empty blocks are omitted. If $E=\emptyset$, use $V_j$ as a single block.
For example, $V_j=\{1,3,5,8\}$ with the usual numerical order and a
single boundary value $5$ gives the blocks $\{1,3\}$ and $\{5,8\}$.
For each attribute $j\in[d]$, choose one block $B_j$ from the partition
of $V_j$. We call the Cartesian product $C=B_1\times\cdots\times B_d$
a \emph{cell}. Use its upper template $\bigvee C$
(Definition~\ref{def:upper_template}) as the candidate for $C$.
Since $C$ is a product of nonempty finite blocks, $\bigvee C\in C$
and $v\preceq\bigvee C$ for every $v\in C$.

For each $e\in E$ and attribute $j$, the truth value of
$\operatorname{pr}_j(e)\preceq_j \operatorname{pr}_j(v)$ is constant as $\operatorname{pr}_j(v)$ varies within a block $B_j$.
Consequently, the set of evaluated templates satisfying $e\preceq v$
is constant on $C$, so~\eqref{eq:optimistic_cost} gives the same
$\underline g_E(v)$ throughout $C$. Thus, for every $v\in C$,
\[
\underline g_E(\bigvee C)=\underline g_E(v),
\qquad \mathrm{Cov}(v)\subseteq\mathrm{Cov}(\bigvee C).
\]
Hence $\bigvee C$ can replace $v$ without changing its optimistic
cost or losing coverage. Let $\mathcal A_E$ be the set of these
candidates $\bigvee C$ after discarding those that cover no sample.

\paragraph{Step 3: Remove redundant candidates}
\begin{lemma}[Pruning by covered samples]
\label{lem:coverage_pruning}
Discard a candidate $r\in\mathcal A_E$ if some $r'\in\mathcal A_E$
satisfies
\[
\underline g_E(r)=\underline g_E(r'),
\qquad \mathrm{Cov}(r)\subsetneq\mathrm{Cov}(r').
\]
Among the remaining candidates with identical optimistic cost and
covered sample set, retain exactly one. The resulting set
$\mathcal R_E$ satisfies Definition~\ref{def:representatives}.
\end{lemma}
\begin{proof}
We first show that every template $v\in S$
with nonempty coverage can be replaced by a candidate in $\mathcal A_E$
without changing its optimistic cost or losing coverage. We then show
that pruning $\mathcal A_E$ retains exactly the maximal pairs in $\mathcal P_E$.

\emph{(i) Replacement by a candidate.}
Fix $v\in S$ with $\mathrm{Cov}(v)\neq\emptyset$.
First, Lemma~\ref{lem:observed_value_restriction} allows us to replace
$v$ by $\widehat v\in G_n$ with the same optimistic cost and covered
sample set. Next, let $C$ be the cell containing $\widehat v$ and
take its upper template $r:=\bigvee C$. By Step~2,
\begin{align*}
\underline g_E(r)&=\underline g_E(\widehat v)=\underline g_E(v),\\
\mathrm{Cov}(v)&=\mathrm{Cov}(\widehat v)\subseteq\mathrm{Cov}(r).
\end{align*}
Since $\mathrm{Cov}(v)\neq\emptyset$, this candidate $r$ is not
discarded in Step~2, so $r\in\mathcal A_E$.

\emph{(ii) Identification of the retained pairs.}
First, for every maximal pair $(c,K)\in\mathcal P_E$, there
exists $r\in\mathcal A_E$ such that $\underline g_E(r)=c$ and
$\mathrm{Cov}(r)=K$. Indeed, applying part~(i) to a
template $v$ realizing $(c,K)$ yields $r\in\mathcal A_E$ with
$\underline g_E(r)=c$ and $K\subseteq\mathrm{Cov}(r)$.
Maximality of $(c,K)$ forces $\mathrm{Cov}(r)=K$.
Moreover, no candidate of cost $c$ can cover a strict superset of $K$,
so this pair survives the pruning.

Conversely, take $r\in\mathcal A_E$ such that
$(\underline g_E(r),\mathrm{Cov}(r))$ is not maximal in $\mathcal P_E$.
Then some $v\in S$ satisfies $\underline g_E(v)=\underline g_E(r)$
and $\mathrm{Cov}(r)\subsetneq\mathrm{Cov}(v)$.
Applying part~(i) to $v$ gives $r'\in\mathcal A_E$ with
$\underline g_E(r')=\underline g_E(v)$ and
$\mathrm{Cov}(v)\subseteq\mathrm{Cov}(r')$.
Thus $\underline g_E(r')=\underline g_E(r)$ and
$\mathrm{Cov}(r)\subsetneq\mathrm{Cov}(r')$, so $r$ is discarded.
Hence the retained pairs are exactly the maximal pairs in
$\mathcal P_E$, and keeping one candidate per pair gives the required
representative set $\mathcal R_E$.
\end{proof}

By Proposition~\ref{prop:milp}, using this $\mathcal R_E$ in the master
\ac{milp} preserves the optimal value.
All three steps use only sample attributes and the costs already
evaluated on $E$; they require no additional evaluations of $g$.

\standalonereferences

\section{Network and Workload Settings}
\label{appendix:experiment_details}

\begin{figure}[t]
\centering
\includegraphics[width=\columnwidth]{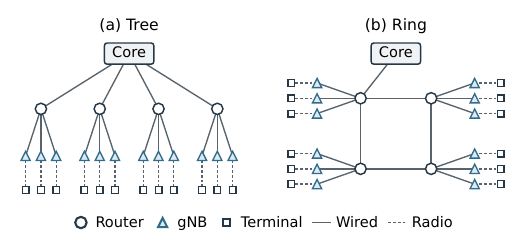}
\caption{Physical downlink topologies. Each router connects to three
\acp{gnb}, each with one terminal. Traffic follows shortest-path distribution
trees; link parameters are listed in
Table~\ref{tab:experiment_network}.}
\label{fig:experiment_topologies}
\end{figure}

The traffic, queueing, and resource-allocation models follow
\cite{kobayashi2026decomposition}. Figure~\ref{fig:experiment_topologies}
shows the two fixed downlink networks, and Table~\ref{tab:experiment_network}
lists their link parameters. Radio capacity is an abstract throughput preset
derived from the maximum-data-rate expression in
TS~38.306~\cite{3gpp38306}. 

\begin{table}[t]
\centering
\caption{Physical links in the fixed downlink networks.}
\label{tab:experiment_network}
\footnotesize
\setlength{\tabcolsep}{3.5pt}
\begin{tabular}{@{}llrrrr@{}}
\toprule
Network & Resource & Count & Gbit/s & km & Queue slots\\
\midrule
Tree & Core--router & 4 & 50 & 30 & 512\\
 & Router--\acs{gnb} & 12 & 20 & 70 & 256\\
 & \acs{gnb}--terminal & 12 & 17.83 & 0 & 128\\
\midrule
Ring & Core--router & 1 & 200 & 30 & 512\\
 & Router--router & 4 & 50 & 30 & 256\\
 & Router--\acs{gnb} & 12 & 20 & 10 & 256\\
 & \acs{gnb}--terminal & 12 & 17.83 & 0 & 128\\
\bottomrule
\end{tabular}
\end{table}

Table~\ref{tab:experiment_requests} defines the experimental service-family
mixture. A family is chosen with probability $1/3$, and its offered rate and
delay threshold are drawn independently and uniformly from the listed
intervals. Arrivals are Poisson, with exponentially distributed
packet sizes of mean 1,600 bits.
\begin{table}[t]
\centering
\caption{Request distribution: equally weighted downlink service families.}
\label{tab:experiment_requests}
\footnotesize
\setlength{\tabcolsep}{4pt}
\begin{tabular}{@{}lrr@{}}
\toprule
Family & Offered rate (Mbit/s/\acs{gnb}) & Delay (ms)\\
\midrule
Remote operation & $[0.5,2]$ & $[3,5]$\\
XR video & $[10,60]$ & $[8,10]$\\
Buffered media & $[30,100]$ & $[100,300]$\\
\bottomrule
\end{tabular}
\end{table}

\standalonereferences

\section{Implementation of the Proposed Method}
\label{appendix:experimental_implementation}

The experiments use Algorithm~\ref{algo:method} with the initialization
and upper-bound refinement described below.
We use HiGHS~1.8.0~\cite{huangfu2018highs} to solve the master and refinement
MILPs. 

\subsection{Initialization of the Proposed Method}
\label{appendix:implementation_initialization}

\subsubsection{Initial catalog}
The proposed method applies L2 $k$-means to $S_n$ and selects each cluster's
upper template~\eqref{eq:upper_template}. The resulting catalog supplies
the initial incumbent, and its evaluated templates enter $E_0$. 

\subsubsection{Lower-bound anchors}
\begin{figure}[t]
\centering
\includegraphics[width=0.72\columnwidth]{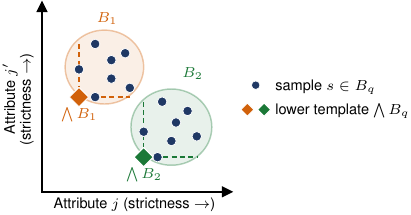}
\caption{Lower-bound strengthening with lower templates.
For every $s\in B_q$, monotonicity gives
$g(\bigwedge B_q)\leq g(s)$. Taking a lower template for each cluster
is expected to yield tighter lower bounds for the points in that cluster.}
\label{fig:lower_templates}
\end{figure}

We strengthen the initial lower estimator with lower templates from a
separate $k$-means clustering of $S_n$
into ten clusters.
Analogously to the upper template~\eqref{eq:upper_template}, define the
\emph{lower template} of a nonempty finite set $T\subseteq S$ by taking
the least stringent requirement of each attribute:
\begin{align}
\label{eq:lower_template}
\bigwedge T\!:=\!\left(\!
\min_{\preceq_1}\{\operatorname{pr}_1(s)\!:\!s\!\in\! T\},\ldots,
\min_{\preceq_d}\{\operatorname{pr}_d(s)\!:\!s\!\in\! T\}
\!\right).
\end{align}
Since $\bigwedge T\preceq s$ for every $s\in T$, monotonicity gives
$g(\bigwedge T)\leq g(s)$.
Let $B_1,\ldots,B_{10}$ denote these ten clusters and define
\begin{align}
E_{\min}&:=\{\bigwedge B_q:q\in[10]\}.
\label{eq:implementation_lower_templates}
\end{align}
We evaluate $g$ at these lower templates before the first master solve
and use their costs as lower bounds.
Figure~\ref{fig:lower_templates} illustrates this construction for two clusters.

These anchors also enter $E_0$. Since $\underline g_E$ uses all of $E$,
each anchor raises $\underline g_E$ at every template that dominates it.
All distinct initial-catalog and lower-template evaluations
count toward the evaluation budget.

\subsection{Upper-Bound Improvement Using Evaluated Templates}
\label{appendix:implementation_upper_bound}

The experimental implementation strengthens the upper-bound update in
line~\ref{line:method_incumbent_update} of Algorithm~\ref{algo:method}
through the following two refinements. Neither requires additional
evaluations of $g$.

\subsubsection{Reassignment within the current catalog}
\label{appendix:implementation_reassignment}
The master selects $\bar y$ using the optimistic cost model $\underline g_E$
(line~\ref{line:method_master}). Once the subproblem evaluates the templates
in $J(\bar y)$ (line~\ref{line:method_subproblem}), this assignment need not
be optimal under $g$, even for the same catalog.
We therefore keep $J(\bar y)$ fixed and reassign each sample to its cheapest
eligible template, or to the outside option if none is eligible. This yields
$f_n(J(\bar y))\leq \phi_g(\bar y)$ without an additional MILP.
We compare this cost with that of the previous
incumbent and retain the better catalog. 

\subsubsection{Catalog reselection from evaluated templates}
\label{appendix:implementation_catalog_reselection}
The subproblem optimizes templates for the fixed assignment $\bar y$
(line~\ref{line:method_subproblem}). Templates evaluated in different
iterations may form a cheaper catalog when combined, with samples
reassigned accordingly. We therefore jointly choose a size-$k$ catalog
from $E$ and its sample assignments using the known costs $g(e)$.
For $|E|\geq k$, define
\begin{align}
U_E\in\arg\min_{\substack{U\subseteq E\\|U|=k}}f_n(U).
\label{eq:evaluated_catalog_selection}
\end{align}
We solve this finite catalog-selection problem as an additional MILP.
We adopt a feasible catalog from
\eqref{eq:evaluated_catalog_selection} only if its training cost is lower
than the incumbent's cost after reassignment. 

The additional MILP is solved once after initialization and whenever an
oracle batch adds evaluated templates. Each solve has a 30-second limit,
capped by the remaining algorithm-time budget, and retains the incumbent
if no improvement is found. Its model construction and solution time are
included in $T_{\rm alg}$. 
\standalonereferences

\ifcitationseen
  \bibliographystyle{IEEEtran}
  \bibliography{references}

\begin{thebibliography}{10}
\providecommand{\url}[1]{#1}
\csname url@samestyle\endcsname
\providecommand{\newblock}{\relax}
\providecommand{\bibinfo}[2]{#2}
\providecommand{\BIBentrySTDinterwordspacing}{\spaceskip=0pt\relax}
\providecommand{\BIBentryALTinterwordstretchfactor}{4}
\providecommand{\BIBentryALTinterwordspacing}{\spaceskip=\fontdimen2\font plus
\BIBentryALTinterwordstretchfactor\fontdimen3\font minus
  \fontdimen4\font\relax}
\providecommand{\BIBforeignlanguage}[2]{{%
\expandafter\ifx\csname l@#1\endcsname\relax
\typeout{** WARNING: IEEEtran.bst: No hyphenation pattern has been}%
\typeout{** loaded for the language `#1'. Using the pattern for}%
\typeout{** the default language instead.}%
\else
\language=\csname l@#1\endcsname
\fi
#2}}
\providecommand{\BIBdecl}{\relax}
\BIBdecl

\bibitem{generic_slice_template}
\BIBentryALTinterwordspacing
{GSMA}, ``Generic network slice template version 10.0,'' Oct. 2024, official
  document NG.116. [Online]. Available:
  \url{https://www.gsma.com/newsroom/wp-content/uploads/NG.116-v10.0-1.pdf}
\BIBentrySTDinterwordspacing

\bibitem{3gpp28530}
{3GPP}, ``Management and orchestration; concepts, use cases and requirements,''
  3GPP, TS 28.530, Version 19.0.0 (Release 19), 2025.

\bibitem{samsung2020network_slicing}
\BIBentryALTinterwordspacing
{Samsung Electronics}, ``Network slicing,'' 2020. [Online]. Available:
  \url{https://www.samsung.com/global/business/networks/insights/white-papers/network-slicing/}
\BIBentrySTDinterwordspacing

\bibitem{3gpp28531}
{3GPP}, ``Management and orchestration; provisioning,'' 3GPP, TS 28.531,
  Version 20.1.0 (Release 20), 2026.

\bibitem{5gamericas2022commercializing}
\BIBentryALTinterwordspacing
{5G Americas}, ``Commercializing {5G} network slicing,'' 2022. [Online].
  Available:
  \url{https://www.5gamericas.org/wp-content/uploads/2022/07/Commercializing-5G-Network-Slicing-Jul-2022.pdf}
\BIBentrySTDinterwordspacing

\bibitem{cisco_cwnc7swg}
\BIBentryALTinterwordspacing
{Cisco Systems, Inc.}, \emph{Cisco Crosswork Network Controller 7.0 Solution
  Workflow Guide}, Mar. 2025. [Online]. Available:
  \url{https://www.cisco.com/c/en/us/td/docs/cloud-systems-management/crosswork-network-controller/7-0/Solution-Workflow-Guide/bk-crosswork-network-controller-7-0-solution-workflow-guide.pdf}
\BIBentrySTDinterwordspacing

\bibitem{vmware_tca_slice_catalog_32}
\BIBentryALTinterwordspacing
{VMware by Broadcom}. (2025) Managing network slice catalog. {VMware Telco
  Cloud Automation} User Guide, Version 3.2. [Online]. Available:
  \url{https://techdocs.broadcom.com/us/en/vmware-sde/telco-cloud/vmware-telco-cloud-automation/3-2/vmware-telco-cloud-automation-user-guide/managing-network-slice-catalog.html}
\BIBentrySTDinterwordspacing

\bibitem{otieno2024review}
H.~O. Otieno, B.~Malila, and J.~Mwangama, ``Deployment and management of
  intelligent end-to-end network slicing in {5G} and beyond {5G} networks: A
  systematic review,'' \emph{IEEE Access}, vol.~12, pp. 190\,411--190\,433,
  2024.

\bibitem{min2024}
Z.~Min, S.~Gokhale, S.~Shekhar, C.~Mahmoudi, Z.~Kang, Y.~Barve, and A.~Gokhale,
  ``Enhancing {5G} network slicing for {IoT} traffic with a novel clustering
  framework,'' \emph{Pervasive Mob. Comput.}, vol. 104, 2024, {Art.} no.
  101974.

\bibitem{kleywegt2002saa}
A.~J. Kleywegt, A.~Shapiro, and T.~{Homem-de-Mello}, ``The sample average
  approximation method for stochastic discrete optimization,'' \emph{SIAM J.
  Optim.}, vol.~12, no.~2, pp. 479--502, 2002.

\bibitem{etsi_zsm_003}
{ETSI ISG ZSM}, ``Zero-touch network and service management ({ZSM}); end-to-end
  management and orchestration of network slicing,'' ETSI, Group Specification
  GS ZSM 003 V1.1.1, Jun. 2021.

\bibitem{wu2026yang}
\BIBentryALTinterwordspacing
B.~Wu, D.~Dhody, R.~Rokui, T.~Saad, and J.~Mullooly, ``A {YANG} data model for
  the {RFC} 9543 network slice service,'' IETF Internet-Draft
  draft-ietf-teas-ietf-network-slice-nbi-yang-26, May 2026, work in progress.
  [Online]. Available:
  \url{https://datatracker.ietf.org/doc/draft-ietf-teas-ietf-network-slice-nbi-yang/}
\BIBentrySTDinterwordspacing

\bibitem{su2019resource_allocation}
R.~Su, D.~Zhang, R.~Venkatesan, Z.~Gong, C.~Li, F.~Ding, F.~Jiang, and Z.~Zhu,
  ``Resource allocation for network slicing in {5G} telecommunication networks:
  A survey of principles and models,'' \emph{IEEE Netw.}, vol.~33, no.~6, pp.
  172--179, 2019.

\bibitem{otieno2023}
H.~O. Otieno, J.~Mwangama, and B.~Malila, ``A strategy to develop and translate
  custom {5G} network slice templates using machine learning techniques,'' in
  \emph{Proc. IEEE AFRICON}, 2023, pp. 1--6.

\bibitem{Lloyd1982}
S.~P. Lloyd, ``Least squares quantization in {PCM},'' \emph{IEEE Trans. Inf.
  Theory}, vol.~28, no.~2, pp. 129--137, Mar. 1982.

\bibitem{hakimi1965median}
S.~L. Hakimi, ``Optimum distribution of switching centers in a communication
  network and some related graph theoretic problems,'' \emph{Oper. Res.},
  vol.~13, no.~3, pp. 462--475, 1965.

\bibitem{nagarajan2015container}
V.~Nagarajan, K.~K. Sarpatwar, B.~Schieber, H.~Shachnai, and J.~L. Wolf, ``The
  container selection problem,'' in \emph{Proc. Approximation, Randomization,
  and Combinatorial Optimization. Algorithms and Techniques (APPROX/RANDOM)},
  ser. LIPIcs, vol.~40, 2015, pp. 416--434.

\bibitem{hooker2003lbbd}
J.~N. Hooker and G.~Ottosson, ``Logic-based {Benders} decomposition,''
  \emph{Math. Program.}, vol.~96, no.~1, pp. 33--60, 2003.

\bibitem{forbes2024simulation}
M.~A. Forbes, M.~G. Harris, H.~M. Jansen, F.~A. van~der Schoot, and T.~Taimre,
  ``Combining optimisation and simulation using logic-based {Benders}
  decomposition,'' \emph{Eur. J. Oper. Res.}, vol. 312, no.~3, pp. 840--854,
  2024.

\bibitem{baptista2018bocs}
R.~Baptista and M.~Poloczek, ``{Bayesian} optimization of combinatorial
  structures,'' in \emph{Proc. Int. Conf. Mach. Learn. (ICML)}, vol.~80, 2018,
  pp. 462--471.

\bibitem{Kim2024SAA}
S.~Kim, R.~Pasupathy, and S.~G. Henderson, ``A guide to sample average
  approximation,'' in \emph{Handbook of Simulation Optimization}, ser.
  International Series in Operations Research \& Management Science, M.~C. Fu,
  Ed.\hskip 1em plus 0.5em minus 0.4em\relax Springer, 2015, vol. 216, pp.
  207--243.

\bibitem{kobayashi2026decomposition}
M.~Kobayashi, A.~Suzuki, R.~Kawahara, and M.~Kobayashi, ``Contextual
  bandit-based decomposition of network slice requirements under cumulative
  resource budget constraints,'' 2026, arXiv:2609.09624.

\bibitem{3gpp38306}
{3GPP}, ``{NR}; user equipment ({UE}) radio access capabilities,'' 3GPP, TS
  38.306, Version 18.6.0 (Release 18), 2025.

\bibitem{huangfu2018highs}
Q.~Huangfu and J.~A.~J. Hall, ``Parallelizing the dual revised simplex
  method,'' \emph{Math. Program. Comput.}, vol.~10, no.~1, pp. 119--142, 2018.

\end{thebibliography}
\fi

\end{document}